\documentclass[lettersize,journal]{IEEEtran}
\usepackage{amsmath,amsfonts}
\usepackage{algorithmic}
\usepackage{algorithm}
\usepackage{array}
\usepackage[caption=false,font=footnotesize]{subfig}
\usepackage{textcomp}
\usepackage{gensymb}
\usepackage{stfloats}
\usepackage{url}
\usepackage{verbatim}
\usepackage{graphicx}
\usepackage{cite}
\usepackage{color}
\usepackage{amsmath}
\usepackage{multirow}
\usepackage{tabularx}
\usepackage{ragged2e}
\usepackage[normalem]{ulem}
\usepackage{steinmetz}
\usepackage{titlesec}
\titlespacing*{\subsection}
{0pt}    
{0.5\baselineskip} 
{0.2\baselineskip}  
\begin{document}
\title{A Single-Inductor $N$:1 Resonant Switched-Capacitor Ladder Converter with Efficient Continuous Voltage Conversion Capability}

\author{\IEEEauthorblockN{Sandeep Reddy Kukunuru, M Mahmudul Hasan Sajeeb, Rohit Reddy Karnaty, and Loai G. Salem} \\
\thanks{The authors are with the Department of Electrical and Computer Engineering, University of California at Santa Barbara, Goleta, CA 93106 USA (e-mail: sandeepreddy@ucsb.edu; msajeeb@ucsb.edu; rkarnaty@ucsb.edu;  salem@ucsb.edu).}
}



\maketitle

\begin{abstract}
This paper presents an $N$:1 resonant switched-capacitor (ReSC) ladder converter of an output voltage that can be regulated continuously below the native $N$:1 resonant ratio while maintaining high efficiency. The topology can be generalized to any $N$-to-1 dc-dc conversion application while requiring only a single inductor. Efficient wide-range voltage regulation is realized by operating the bottom stack of the ReSC switches as diodes to maintain their switching always in phase with the inductor current, and hence, circumvent circulating currents that increase the inductor RMS current and ohmic losses under frequency control. A theoretical model has been developed to predict the output voltage and efficiency of the converter, with simulation results confirming its accuracy for $N$ = 1, 2, and 4. A hardware prototype on a printed circuit board is developed to validate the performance of the proposed frequency-controlled regulation scheme.
The prototype achieves a peak efficiency of 97\% (94.3\%) with a load of 30\,$\Omega$ (resp. 6\,$\Omega$) and an efficiency $>$90\% while generating a 9\,V USB-compatible voltage from a 40\,V-to-60\,V input. The measurement results match the model predictions within 0.63\% accuracy.
\end{abstract}

\begin{IEEEkeywords}
dc-dc converter, data center power, resonant SC converter, ladder SC converter.
\end{IEEEkeywords}

\section{Introduction}
\IEEEPARstart{T}{he} rapid advancement  of cloud computing, artificial intelligence, and high-performance computing has multiplied the power consumption of data centers. The overall energy consumption of data centers accounts for approximately 1$-$2\% of the total global electricity demand, with estimates of more than 200 terawatt hours (TWh) per year \cite{datacentreref}. To mitigate energy-related costs and environmental impact, modern data centers have transitioned from traditional 12\,V architectures to 48\,V power distribution networks, which reduces power distribution conduction losses and improves efficiency \cite{48Vrailref}. However, efficient steep voltage conversion from the 48\,V bus to lower voltages, such as 9\,V, remains a critical challenge. Improved power efficiency is not only required to save energy but also to reduce the cost of thermal cooling systems, which are used to manage heat dissipation in high-power density data centers. 

Resonant dc-dc converters have gained widespread use in many applications~\cite{rsc_Mierczuk, b1, rsc_ZYe1, rsc_ZYe2, b13, b14, rsc_ZYe4, b15, b187, rsc_Deshmukh, rsc_Mudiyanselage, rsc_TICLLLC, rsc_Wolfspeed1, rsc_Wolfspeed2, rsc_IOLee, rsc_BKLee}. Resonant dc-dc conversion differs from pulse-width modulated (PWM) dc-dc conversion and switched-capacitor (SC) dc-dc converters in that it handles power in a sinusoidal form, which facilitates soft device switching and allows a significant decrease in switching losses and noise~\cite{b23, mohan, kassakian, lidow, ReSCbook}. A subclass of resonant dc-dc converters is established by incorporating an inductor or multiple inductors in an existing SC dc-dc converter topology to soft charge and discharge all or some of the employed flying capacitors \cite{b16, b17, b18, b19, b20, b11, b26, ISCAS_conf, scrsc_Salem, esserc_San, b0, b5, spvcicc, spvjssc, lsccicc, sscl, mwtlms, iscasms, iscas2, thesis}.

On the other hand, it was proven that the SC ladder topology meets the fundamental limit for the minimum possible switch V-A stress applicable to all dc-dc converters (derived by Wolaver in 1969 using graph theory \cite{Wolaver}) \cite{seemanCOMPEL}, and hence, the SC ladder topology produces the smallest possible conduction loss compared to the remaining SC and switched-inductor (SL) dc-dc topologies for the same conversion ratio, using the same FET technology, and under identical output powers~\cite{seemanladdersc, b24, b25, b2, b3, b4}. Prior work combined the benefits of the SC ladder topology and resonant power conversion to realize efficient 1:$N$ step-up voltage conversion~\cite{b11}. Although the approach in~\cite{b26} demonstrated the resonant ladder advantages for $N$:1 voltage conversion, efficiency is not fully optimized for voltages lower than $V_{in}$/$N$. This inefficiency results from extra conduction losses due to additional circulating currents when the switching frequency is varied to regulate the output voltage below $V_{in}$/$N$. To address this issue and enable efficient steep voltage conversion, this work presents an $N$:1 resonant SC ladder dc-dc converter with diode-operated bottom switches to realize the required ac-to-dc conversion and make the voltage switching of the bottom half bridge in-phase with the resonant inductor current.
To provide insight into the operation of the proposed $N$:1 resonant SC ladder dc-dc converter and evaluate its performance, a theoretical model is developed to predict the steady-state conversion ratio and efficiency. 

This article is organized as follows. Section II reviews voltage regulation techniques used in prior $N$:1 resonant SC converters and highlights their limitations. Section III introduces the proposed diode-operated switches that enable automatic phase-shift optimization. Section IV presents the steady-state modeling and analysis of both the proposed solution and prior approaches. Finally, Section V describes the prototype implementation while Section VI discusses the experimental results.
\section{Prior $N$:1 Resonant SC Converters Voltage Regulation Techniques}

\begin{figure}[!tb]
\centering
\includegraphics[width=2.0in]{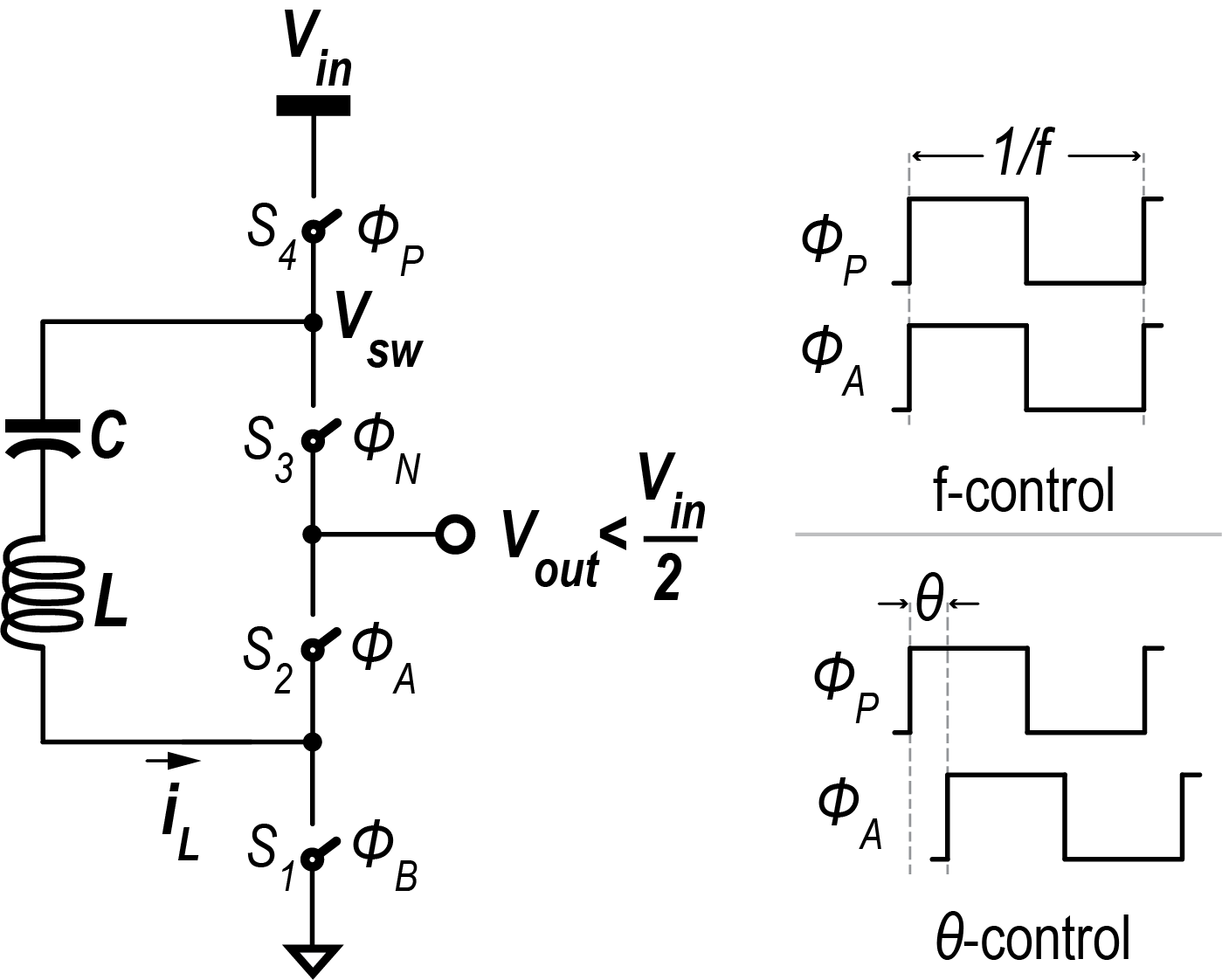}
\caption{Prior 2:1 ReSC $V_{out}$ regulation using frequency and phase-shift controlled clocks.}
\label{fig.prior_vout_control}
\end{figure}
A 2:1 resonant SC converter constructed with two half bridges is shown in Fig.~\ref{fig.prior_vout_control}~\cite{b16}. The two half bridges are driven by complementary clocks: $\phi_A$/$\phi_B$ and $\phi_P$/$\phi_N$. To minimize MOSFET switching losses, the converter is operated in the zero voltage switching (ZVS) region by switching the converter at a frequency, $f$, slightly higher than the resonant tank frequency, $f_{o}$. In this way, the LC tank introduces an inductive impedance, making the switch current lag its voltage. The lossless output voltage, $V_{out}$, regulation around the resonant SC (ReSC) native ratio can be achieved via phase-shift control ($\theta$-control in Fig.~\ref{fig.prior_vout_control}, bottom right) or switching frequency tuning ($f$-control in Fig.~\ref{fig.prior_vout_control}, top right)~\cite{b17, b18, b19, b20}. In phase-shift control \cite{b17}, the phase difference, $\theta$, between the clocks of the two stacked half-bridge inverters $S_{1}$/$S_{2}$ and $S_{3}$/$S_{4}$ is varied to regulate $V_{out}$. In frequency modulation techniques, the switching frequency, $f$, is varied around $f_o$ to regulate $V_{out}$ at the desired target value, while the phase shift between the clocks of the two half bridges is kept constant~\cite{b19}. In frequency control techniques, the phase angle, $\theta$, is not dynamically changed with $f$, but rather, its value is typically selected to realize a desired output voltage range.

In a 2:1 ReSC, the amplitude of the inductor current $i_{L}$ is given by $V_{sw,1}/(Z_{tank} + Z_e)$, where $Z_{e}$ is the equivalent impedance of the dc load, $R_{load}$, that is connected to $V_{out}$, as seen by the tank, and $V_{sw,1}$ is the fundamental component in the square voltage generated by the top switching node, $V_{sw}$, shown in Fig.~\ref{fig.prior_vout_control}.
Both control techniques result in a phase difference between the inductor current and the bottom half bridge switching node voltage, introducing an imaginary part in $Z_{e}$. The imaginary part of $Z_{e}$ initiates a reactive power component and circulating currents that increase the inductor RMS current and the resulting ohmic loss. To address this issue, this paper proposes to operate the bottom two switches as diodes, as shown in Fig.~\ref{fig.proposed_solution} (right), similar to a classical series-resonant dc-dc converter~\cite{b23, b21, b22}, which is shown in Fig.~\ref{fig.proposed_solution} (left), to ensure that $V_{sense}$ and the inductor sinusoidal current are always in phase (i.e., $\theta=\phi$, where $\phi$ is the phase angle of inductor current) under $f$-control, making $Z_{e}$ always feature a real value with a zero imaginary part.

\section{Proposed Diode-Operated Switches for Automatic Optimal Phase-Shift Tracking}
\begin{figure}[!tb]
\centering
\includegraphics[width=3.4in]{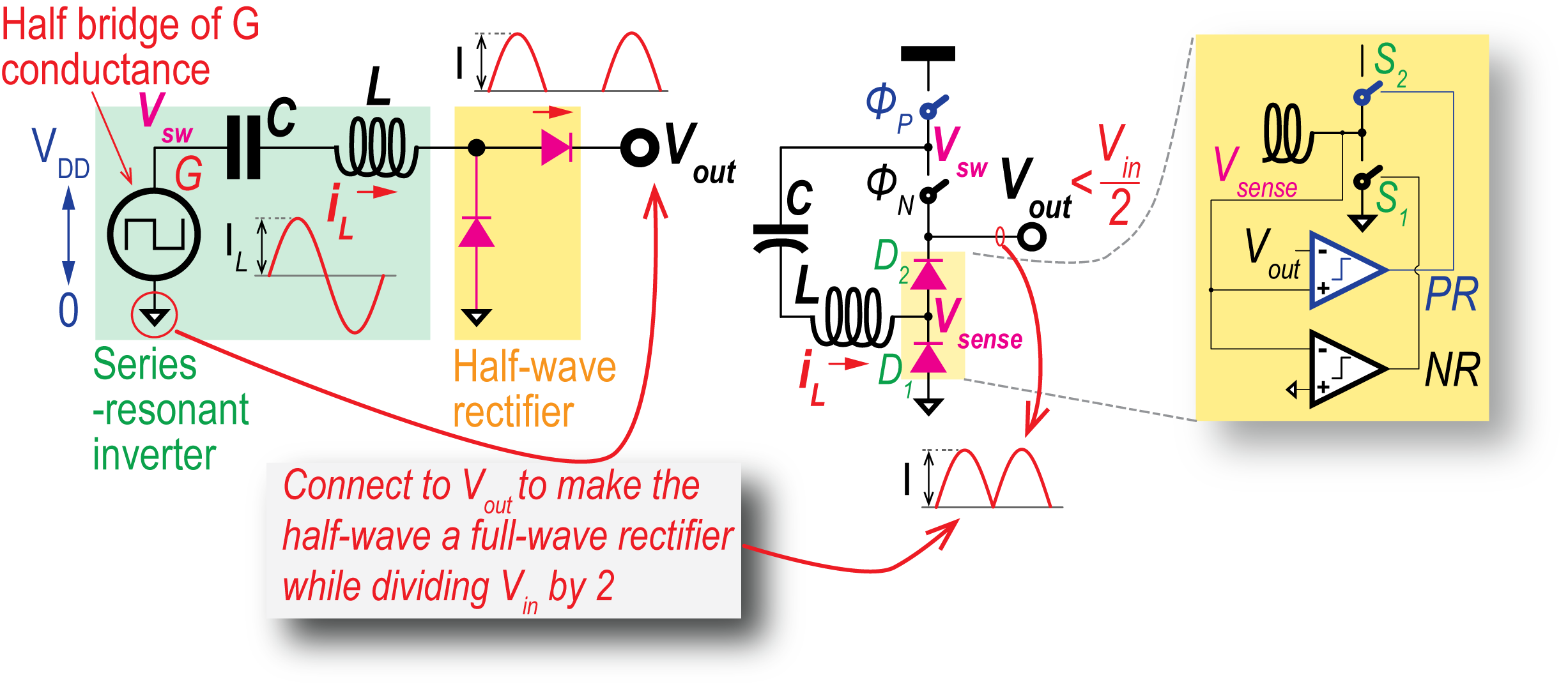}
\caption{Proposed diode-operated switches in a 2:1 ReSC converter (right); a 2:1 ReSC can be considered as a stacked version of a traditional series-resonant dc-dc converter.}
\label{fig.proposed_solution}
\end{figure}

In a traditional series-resonant dc-dc converter, as shown in Fig.~\ref{fig.proposed_solution} (left), a half-bridge inverter generates a square wave, $V_{sw}$, whose fundamental frequency component, $V_{sw1}$, is extracted by a series LC resonant tank, while its higher-order harmonics are filtered. A half-wave rectifier and a capacitor filter at the output convert the generated ac to a dc voltage. Output voltage, $V_{out}$, regulation can be realized by modulating the converter switching frequency around $f_o$~\cite{b23}. As shown in Fig.~\ref{fig.proposed_solution}, a resonant 2:1 SC converter can be considered as a traditional series-resonant dc-dc converter with its half-bridge inverter ground terminal connected to $V_{out}$, i.e., the half-bridge inverter is stacked on top of the half-wave rectifier. Stacking the half-bridge inverter on top of the half-wave rectifier features two advantages: 1) the inverter voltage swing is divided by two, $V_{in}$/2, and 2) the stack allows the negative halves of the resonant tank current to reach $V_{out}$ via the third switch, $S_{3}$, shown in Fig.~\ref{fig.prior_vout_control}, and hence, the original half-wave rectifier behaves as a full-wave rectifier despite using only two diodes. 

\begin{figure}[!tb]
\centering
\includegraphics[width=2.2in]{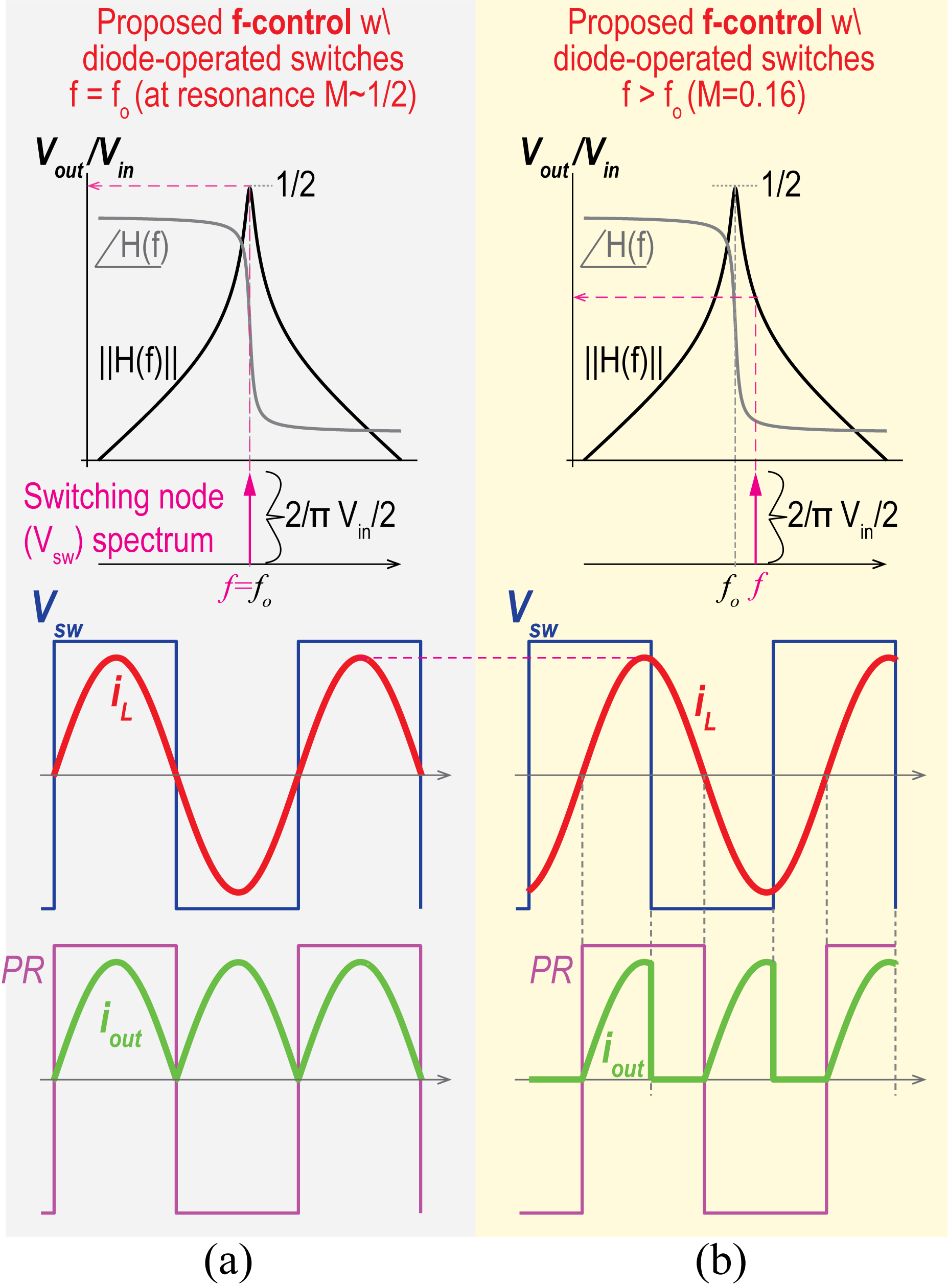}
\caption{2:1 ReSC key voltage and current waveforms under the proposed f-control using diodes for (a) $f = f_{o}$ and (b) $f>f_{o}$.}
\label{fig.res_control}
\end{figure}

In this paper, efficient $V_{out}$ regulation below the native 2:1 SC resonant converter output, $V_{in}$/2, is achieved by increasing the half-bridge inverter switching frequency, $f$, above the tank resonant frequency, $f_{o}$, while operating the MOSFET switches $S_{1,2}$ as diodes~\cite{ISCAS_conf}. The gate control signals of the bottom switches $S_{1,2}$ in a 2:1 SC resonant converter are generated by comparing $V_{sense}$ with $V_{out}$ and ground, as illustrated in Fig.~\ref{fig.proposed_solution} (right). In this way, the switching node voltage of the bottom half bridge in a 2:1 SC resonant converter, $V_{sense}$, is kept in phase with the inductor current, $i_L$, under different $f$ values, which eliminates the imaginary part of $Z_{e}$ and minimizes the converter conduction loss. As illustrated in Fig.~\ref{fig.res_control}(a) and (b), when $f$ is increased above $f_{o}$ for $V_{out}$ regulation, the tank impedance becomes more inductive and the phase-lag, $\phi$, of the inductor current, $i_{L}$, behind $V_{sw}$ (the switching node of the stacked inverter in Fig.~\ref{fig.prior_vout_control}) increases. Therefore, operating the bottom half-bridge switches $S_{1,2}$ from a clock source of a fixed phase-shift $\theta$ relative to the top half-bridge clock, while varying $f$, as suggested in~\cite{b19}, leads to high circulating currents. That is, the current delivered to $V_{out}$, $i_{out}$ in Fig.~\ref{fig.res_control}(a) and (b), will have negative parts that represent electric charge taken from $V_{out}$ and returned back to the tank circuit when $\theta$ is different from $\phi$, and hence, a larger $i_{L}$ amplitude is needed for the same average load current value. The higher $i_{L}$ amplitude and RMS value for the same load under phase misalignment (i.e., $\theta \neq \phi$) leads to larger ohmic loss that degrades the converter efficiency, $\eta$, rapidly as $V_{out}$ is regulated away from $V_{in}$/2. On the other hand, by operating $S_{1,2}$ as diodes, $\theta$ automatically tracks $\phi$, which eliminates the circulating currents and keeps $\eta$ high as $V_{out}$ is decreased.

\begin{figure}[!tb]
\centering
\includegraphics[width=3.4in]{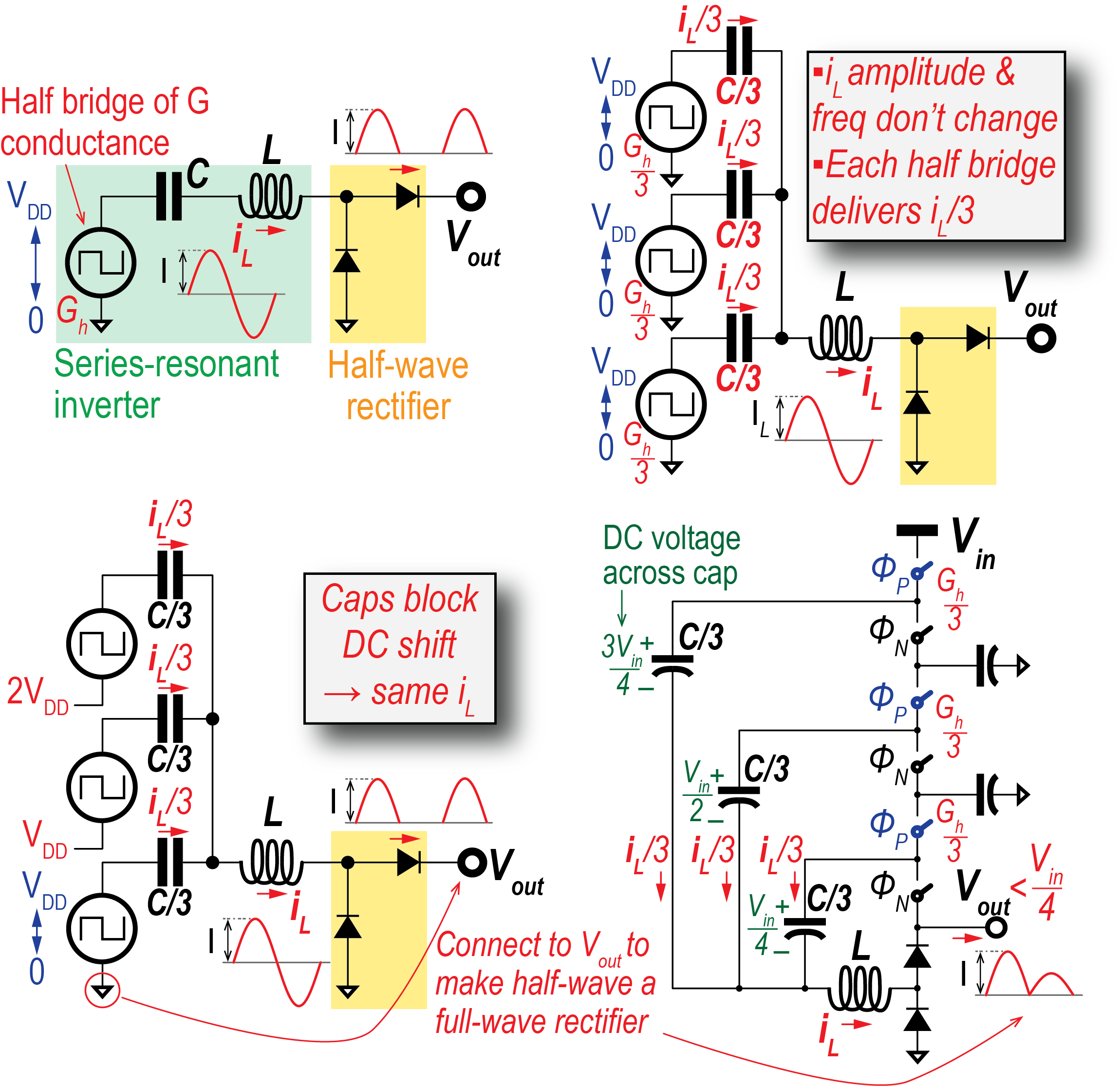}
\caption{Proposed $N$:1 ReSC ladder topology derivation from a series-resonant dc-dc converter. The series-resonant inverter (top left) is split ($N$-1) ways (top right). Then, the ground of the top ($N$ - 1) half bridges are dc shifted to integer multiples of the step voltage, $V_{DD}$ (bottom left). The ground of the bottom half bridge is connected to $V_{out}$ while the remaining half bridges are stacked on top of each other.}
\label{fig.proposed_resc}
\end{figure}
A 1:$N$ ReSC ladder topology for step-up voltage conversion was demonstrated by Yeung~\cite{b11} in 2004, while~\cite{b26} demonstrated its use for step-down conversion. Both topologies did not utilize the proposed diode-operated bottom switches that enable efficient continuous voltage conversion. The proposed $N$:1 ReSC ladder topology with diode-operated bottom switches shown in Fig.~\ref{fig.proposed_resc} (bottom right) enables resonance of all $N-$1 flying capacitors using only a single series inductor, unlike~\cite{b1}, within a 2-phase periodic sequence in contrast to the required multi-phase clocking in~\cite{b13} and~\cite{b14}, all while maintaining a constant $f_{o}$ from one phase to another, unlike~\cite{b13, b14, b15}. In addition, the inductor in the proposed converter is connected with the $N-$1 flying capacitors in a series tank configuration instead of the output-side connection employed in~\cite{b13, b14, b15} that places the switches between the inductor and capacitors, exposing them to high voltage swings.  

As shown in Fig.~\ref{fig.proposed_resc}, the proposed $N$:1 ReSC ladder can be seen analogous to a traditional series-resonant dc-dc converter, with its half-bridge inverter and resonant tank capacitance split into $N-$1 identical parts, each of $G_{h}/(N-1)$ conductance and $C/(N-1)$ capacitance, that are stacked on top of each other with the bottom half-bridge inverter ground connected to $V_{out}$. The $N$:1 ReSC ladder topology depends on four facts as illustrated in Fig.~\ref{fig.proposed_resc}: 1) in a series-resonant circuit, the series capacitance and half-bridge inverter can be split $N-$1 ways without affecting the resonant frequency $f_{o}=1/(2\pi\sqrt{LC})$, since the total capacitance seen by $L$ remains the same; 2) stacking the $N$ half bridges on top of each other does not change the amplitude or frequency of the inductor sinusoidal current than the parallel case, since the capacitors block the dc voltage difference between the stacked domains; 3) connecting the ground-side of the bottom half bridge to $V_{out}$ provides full-wave rectification, increasing the amount of charge delivered to $V_{out}$ in one cycle compared to the traditional case; and 4) the flying ladder automatically balances the voltage across the stacked half-bridge inverters at $(V_{in}-V_{out})/(N-1)$. Note that similar to a 2:1 ReSC, continuous voltage conversion below the native ReSC ladder discrete ratio, 1/$N$, can be realized via frequency control. 

\textcolor{black}{As N is increased, the effect of the full-wave rectification reduces as the current supplied in the second half of the clock cycle reduces. In addition, any parasitic C or L in the LC tank alters the center frequency of the resonant converter, which can be offset by tuning the frequency of operation. A second cell of the resonant converter can be employed and operated at a 180$\degree$ phase shift in order to provide continuous input current \cite{oopresc}.}

\section{Steady-State Sinusoidal Analysis}
\begin{figure}[!tb]
\centering
\includegraphics[width=3.4in]{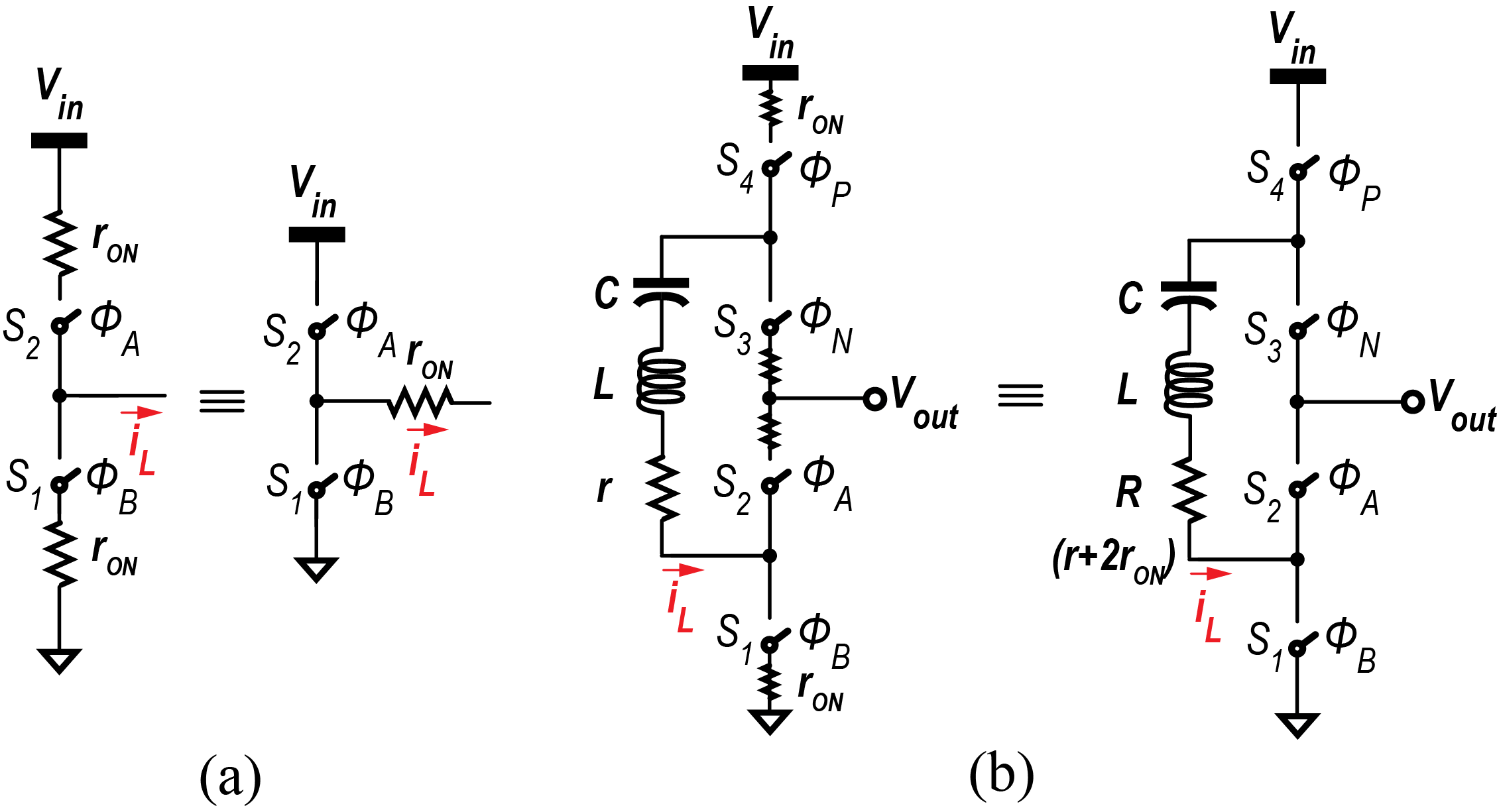}
\caption{Lumping the on-resistance of the MOSFETs in series with the LC tank in (a) a conventional resonant converter and (b) a 2:1 ReSC.}
\label{fig.resistor_modulation}
\end{figure}

As illustrated in Fig.~\ref{fig.resistor_modulation}(a), each MOSFET in the half-bridge inverter of a resonant converter is modeled as an ideal switch in series with an on-resistance, $r_{ON}$. For simplicity, both switches are assumed to have the same $r_{ON}$, while the MOSFET’s intrinsic body-drain pn junction diode is not included in the model. At any instant, one of the two switches is on, and hence, the two $r_{ON}$ can be represented by a single on-resistance in series with the LC tank. Similarly, as shown in Fig.~\ref{fig.resistor_modulation}(b), the on-resistance of the four MOSFETs in a 2:1 ReSC converter can be lumped into a single 2$r_{ON}$ resistance in series with the flying LC tank. Resistance $r$ is the equivalent series resistance (ESR) of the inductor, L, in addition to the ESR of the flying capacitor, C. Therefore, the total conduction loss of the 2:1 ReSC converter is proportional to the square of the tank circuit RMS current as \(I_{L,RMS}^2R\), where $R$ is the total series resistance given by \(2r_{ON}+r\).

\begin{figure}[!tb]
\centering
\includegraphics[width=3.0in]{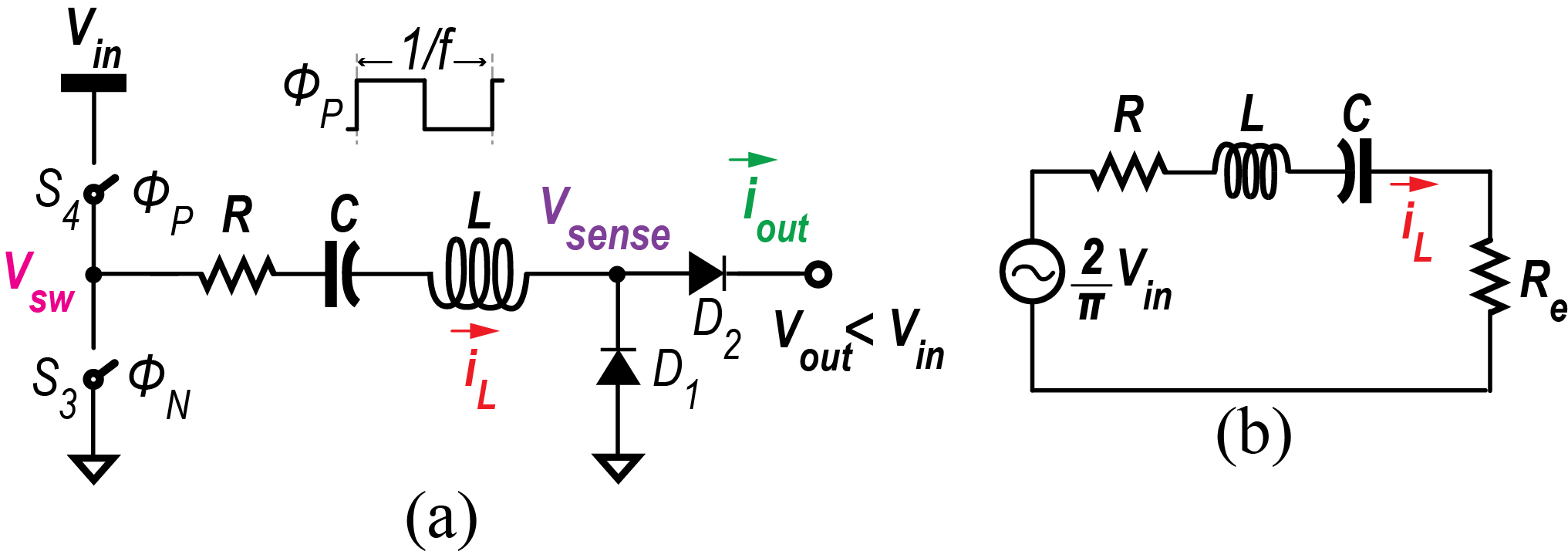}
\caption{(a) The switch diagram of a series-resonant dc-dc converter. (b) Steady-state equivalent circuit of a series-resonant dc-dc converter that models the converter behavior around the fundamental component of $V_{sw}$.}
\label{fig.conv1xresconv}
\end{figure}

\subsection{Series-Resonant dc-dc Converter Loss Analysis}
\label{subsec:conventional_ReSC_loss_analysis}
\begin{figure}[!tb]
\centering
\includegraphics[width=3.4in]{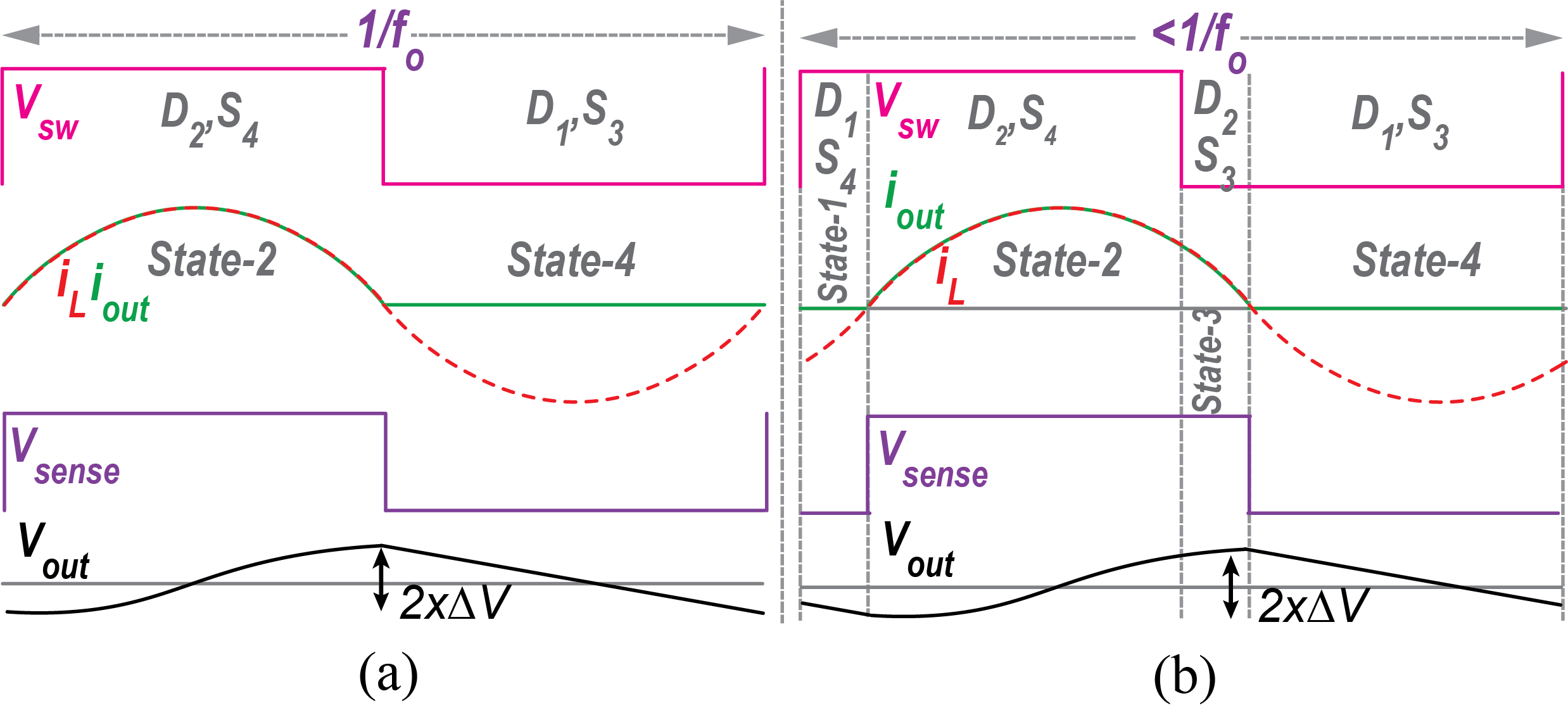}
\caption{Series-resonant converter simulated switching-node voltage, $V_{sw}$, tank current, $i_L$, output current, $i_{out}$, and output voltage, $V_{out}$, (a) at resonance, $f = f_{o}$, and (b) above resonance, $f>f_{o}$.}
\label{fig.timingdiagram_conv1xresconv}
\end{figure}

Under a high quality, $Q$, factor, the input half bridge in the series-resonant dc-dc converter shown in Fig.~\ref{fig.conv1xresconv}(a) can be modeled as a sinusoidal voltage source with an amplitude of $2V_{in}/\pi$. In this case, as illustrated in Fig.~\ref{fig.timingdiagram_conv1xresconv}(a), the input voltage of the rectifier $V_{sense}(t)$ (in Fig.~\ref{fig.conv1xresconv}(a)) is essentially a square wave equal to $V_{out}$ when the tank current, $i_{L}(t)$, is positive and zero when $i_{L}(t)$ is negative, i.e., $V_{sense}(t)$ is in phase with $i_{L}(t)$. Therefore, in steady state and under a high $Q$ factor, the nearly sinusoidal tank current can be expressed as
{%
\setlength{\abovedisplayskip}{0.5\baselineskip}%
\setlength{\belowdisplayskip}{0.5\baselineskip}%
\begin{equation}
\begin{aligned}
i_{L}(t) &= \frac{2}{\pi |Z_o|} \left[ V_{in} \sin\left(\omega t - \theta \right) - V_{out} \sin\left(\omega t - \phi - \theta \right) \right], 
\label{eq:secondmethod}
\end{aligned}
\end{equation}
}
where $Z_o$ is $R + j (\omega L - \frac{1}{\omega C} )$, which denotes the RLC tank's impedance and $\theta$ is its phase, i.e., $\tan^{-1}[(\omega L - \frac{1}{\omega C} )/R]$, while $\phi$ is the phase lag between the tank current and the input half-bridge switching clock.

The output half-wave rectifier can also be modeled by an equivalent impedance, $Z_e = R_e + j X_e$, rather than a sinusoidal voltage source as in \eqref{eq:secondmethod}. Therefore, the nearly sinusoidal tank current, $i_{L}(t)$, can also be expressed as
{%
\setlength{\abovedisplayskip}{0.5\baselineskip}%
\setlength{\belowdisplayskip}{0.5\baselineskip}%
\begin{equation}
\begin{aligned}
i_{L}(t) &= \frac{2V_{in}}{\pi |Z|}\sin\left(\omega t - \phi \right), 
\label{eq:1xconveq}
\end{aligned}
\end{equation}
}
where $|Z|$ is the magnitude of the total impedance, $Z$, that is seen by the half-bridge inverter, i.e., $Z_o + Z_e$, while its phase is $\phi$, which represents the phase lag between the tank current and the input half-bridge switching clock. Note, $Z$ is given by
{%
\setlength{\abovedisplayskip}{0.5\baselineskip}%
\setlength{\belowdisplayskip}{0.5\baselineskip}%
\begin{equation}
\begin{aligned}
|Z| &= \sqrt{\left(R+R_{e}\right)^2 + \left(\omega L - \frac{1}{\omega C}+X_e\right)^2},\\
\phi &= \tan^{-1} \left( \frac{\omega L - \frac{1}{\omega C} + X_e}{R+R_{e}} \right).
\label{eq:Z_eq} 
\end{aligned}
\end{equation}
}
In the following, both relations~\eqref{eq:secondmethod} and~\eqref{eq:1xconveq} for the tank current are used to find two expressions for the conversion ratio, $M$, of the series-resonant dc-dc converter shown in Fig.~\ref{fig.conv1xresconv}(a). As will be shown, the derived expressions have identical values. Note that, using the two approaches not only helps validate the derived expressions but also helps gain insight into the resonant operation, which is key for the modeling of an $N$:1 ReSC ladder converter.
The dc output voltage, $V_{out}$, of the series-resonant converter shown in Fig.~\ref{fig.conv1xresconv}(a) is given by the average of the rectifier's output current, $i_{out}$, times the dc load resistance, $R_{load}$. 
The dc output voltage is thus given by
{%
\setlength{\abovedisplayskip}{0.5\baselineskip}%
\setlength{\belowdisplayskip}{0.5\baselineskip}%
\begin{equation}
V_{out}=\frac{|i_{L}|R_{load}}{\pi},
\label{eq:voutconvseries}
\end {equation}
}
where $|i_L|$ is the amplitude of the tank current, $i_L$. The output voltage is regulated below \(V_{in}\) by increasing the switching frequency, $f$, of the half bridge above \(f_{o}\), which lowers the magnitude of the tank current. 
From~\eqref{eq:secondmethod} and~\eqref{eq:voutconvseries}, the converter dc conversion ratio, $M$, can be found as
{%
\setlength{\abovedisplayskip}{0.5\baselineskip}%
\setlength{\belowdisplayskip}{0.5\baselineskip}%
\begin{equation}
\begin{aligned}
M &= \left[
\frac{1}{\sqrt{1 + T^2}}  \left(
1 + \frac{T \left(1 + \tan\theta \times T\right)}{\tan\theta - T}
\right)
\right]^{-1},\\
T &= \frac{\sin\theta}{\cos\theta + \dfrac{2 R_{load}}{\pi^2 |Z_o|}}.
\end{aligned}
\label{eq:Mconvdiode_secondmethod}
\end{equation}
}

On the other hand, the diode rectifier in Fig.~\ref{fig.conv1xresconv}(a) eliminates the imaginary part, $X_e$, of $Z_e$. The series-resonant dc-dc converter shown in Fig.~\ref{fig.conv1xresconv}(a) thus can be modeled by the equivalent circuit shown in Fig.~\ref{fig.conv1xresconv}(b)~\cite{b23}, where the uncontrolled half-wave rectifier always presents a real load, $R_{e}$, to the tank that is equal to $2 R_{load}/\pi^2$. In this case, from~\eqref{eq:1xconveq} and \eqref{eq:voutconvseries}, $M$ can be found as 
{%
\setlength{\abovedisplayskip}{0.5\baselineskip}%
\setlength{\belowdisplayskip}{0.5\baselineskip}%
\begin {equation}
M = \frac{R_{e}}{|Z|}.
\label{eq:M1x1resc}
\end {equation}
}

As shown in Fig.~\ref {fig.timingdiagram_conv1xresconv}(b), when $f$ is higher than $f_o$, two additional states, state-1 and state-3, appear in addition to the original resonant states, state-2 and state-4 that are shown in Fig. \ref {fig.timingdiagram_conv1xresconv}(a). 
Nevertheless, the relation~\eqref{eq:voutconvseries} between the dc output voltage, $V_{out}$, and the magnitude of the tank current, $i_{L}$, remains the same, because the rectifier input current stays in phase with its input voltage, $V_{sense}$. The additional states only reduce the magnitude of the tank current, $i_{L}$. As a result, the efficiency of the converter, considering the conduction loss only, is independent of the conversion ratio and is constant for a given $R_{load}$
{%
\setlength{\abovedisplayskip}{0.5\baselineskip}%
\setlength{\belowdisplayskip}{0.5\baselineskip}%
\begin{equation}
\eta = \left(1 + \frac{R}{R_{e}}\right)^{-1}.
\label{eq:etaconvseries}
\end {equation}
}

\begin{figure}[!h]
\centering
\includegraphics[width=3.2in]{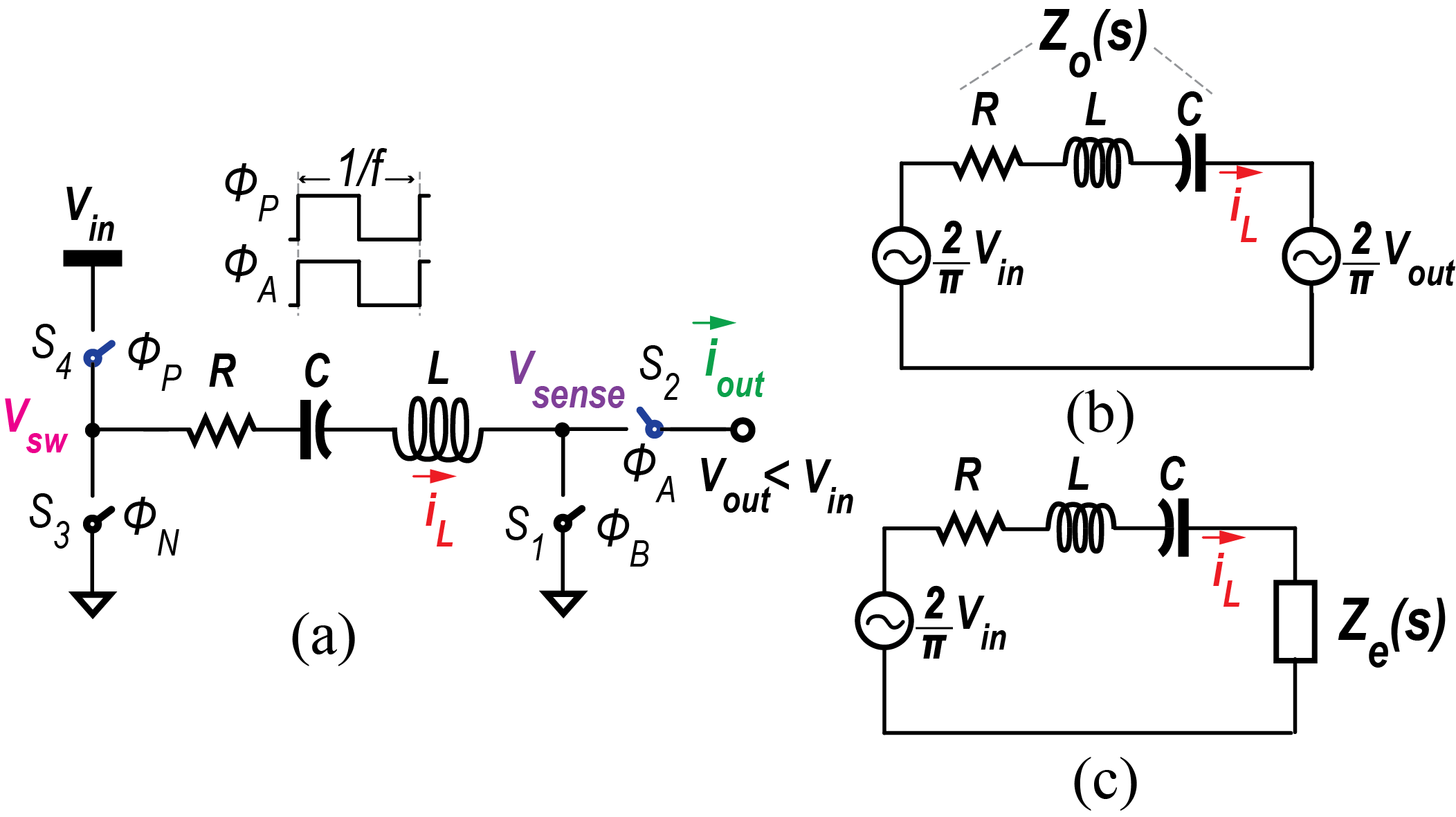}
\caption{(a) The switch diagram of a series-resonant dc-dc converter with a clock-controlled switch rectifier. (b) Steady-state equivalent circuit of the converter that models the behavior of the converter around the fundamental component of $V_{sw}$. (c) Steady-state equivalent circuit of the converter with the rectifier modeled as a reactive impedance.}
\label{fig.1xseries_conv_without_diodes}
\end{figure}

As shown in Fig.~\ref{fig.1xseries_conv_without_diodes}(a), a conventional series-resonant converter can also use a clock-controlled switch rectifier in place of the discussed uncontrolled diode rectifier. The controlled rectifier can be modeled as a sinusoidal source with an amplitude of $2V_{out}/\pi$ that is in phase with the half bridge on the input side, as shown in Fig.~\ref{fig.1xseries_conv_without_diodes}(b). In this case, as $f$ is increased above $f_o$, a phase difference starts to appear between the tank current, $i_L$, and the rectifier's input voltage, $V_{sense}$, and hence, the controlled rectifier presents a reactive impedance, $Z_e(s)$, with an imaginary part to the tank, as shown in Fig.~\ref{fig.1xseries_conv_without_diodes}(c), where $s$ is $j\omega$. The impedance $Z_e$ in polar form is given by
{%
\setlength{\abovedisplayskip}{0.5\baselineskip}%
\setlength{\belowdisplayskip}{0.5\baselineskip}%
\begin{equation} \begin{aligned} 
  &Z_{e}(s) = \frac{2}{\pi^2}R_{load} \cos(\phi) \phase{\phi}.
  \label{eq:zwodiodes}
\end{aligned}\end{equation}
}
The impedance $Z_{e}$ also can be expressed as
{%
\setlength{\abovedisplayskip}{0.5\baselineskip}%
\setlength{\belowdisplayskip}{0.5\baselineskip}%
\begin{equation} \begin{aligned} 
  &Z_{e}(s) = \frac{M}{1-M}Z_o(s) ~ \text{for} ~ M<1.
  \label{eq:zwodiodeszo}
\end{aligned}\end{equation}
} 
Therefore, the phase of $Z_e$, $Z_o$, and $Z$ are the same. 
The tank current can be approximated by~\eqref{eq:secondmethod} with $\phi$ in the second term set to 0, according to the equivalent circuit in Fig.~\ref{fig.1xseries_conv_without_diodes}(b), or~\eqref{eq:1xconveq} according to the model in Fig.~\ref{fig.1xseries_conv_without_diodes}(c). As the phase lag, $\phi$, increases with higher $f$, the magnitude of $Z_e$ in~\eqref{eq:zwodiodes} decreases, which increases the tank current for the same $R_{load}$ and leads to greater conduction loss.

The dc output voltage is given by
{%
\setlength{\abovedisplayskip}{0.5\baselineskip}%
\setlength{\belowdisplayskip}{0.5\baselineskip}%
\begin{equation} \begin{aligned} 
&V_{out}=\frac{|i_{L}|R_{load}}{\pi}\cos(\phi).
\label{eq:phiwodiodes}
\end{aligned}\end{equation}
}
The dc conversion ratio, M, of the converter can be found from~\eqref{eq:phiwodiodes} along with~\eqref{eq:secondmethod} or~\eqref{eq:1xconveq}. Both lead to
{%
\setlength{\abovedisplayskip}{0.5\baselineskip}%
\setlength{\belowdisplayskip}{0.5\baselineskip}%
\begin{equation}
M = \left(1+\frac{R}{R_e(\omega)}\right)^{-1},
\label{eq:Mwodiodes}
\end{equation}
}
where $R_e(\omega)$ is the real part of $Z_e$ which is equal to ~$\frac{2}{\pi^2} R_{load}\cos^2(\phi)$.
The efficiency of the converter is given by
{%
\setlength{\abovedisplayskip}{0.5\baselineskip}%
\setlength{\belowdisplayskip}{0.5\baselineskip}%
\begin{equation}
\eta = \left(1 + \frac{R}{R_e(\omega)} \right)^{-1} = M.
\label{eq:eta1xvseries_withoutdiodes}
\end{equation}
}
It can be observed from~\eqref{eq:Mwodiodes} and~\eqref{eq:eta1xvseries_withoutdiodes} that the efficiency is no longer independent of the conversion ratio. In fact, the efficiency decreases as $f$ is made larger than $f_o$, and hence with higher $\phi$, where the tank current increases for the same $R_{load}$. 

\begin{figure}[!h]
\centering
\includegraphics[width=3.4in]{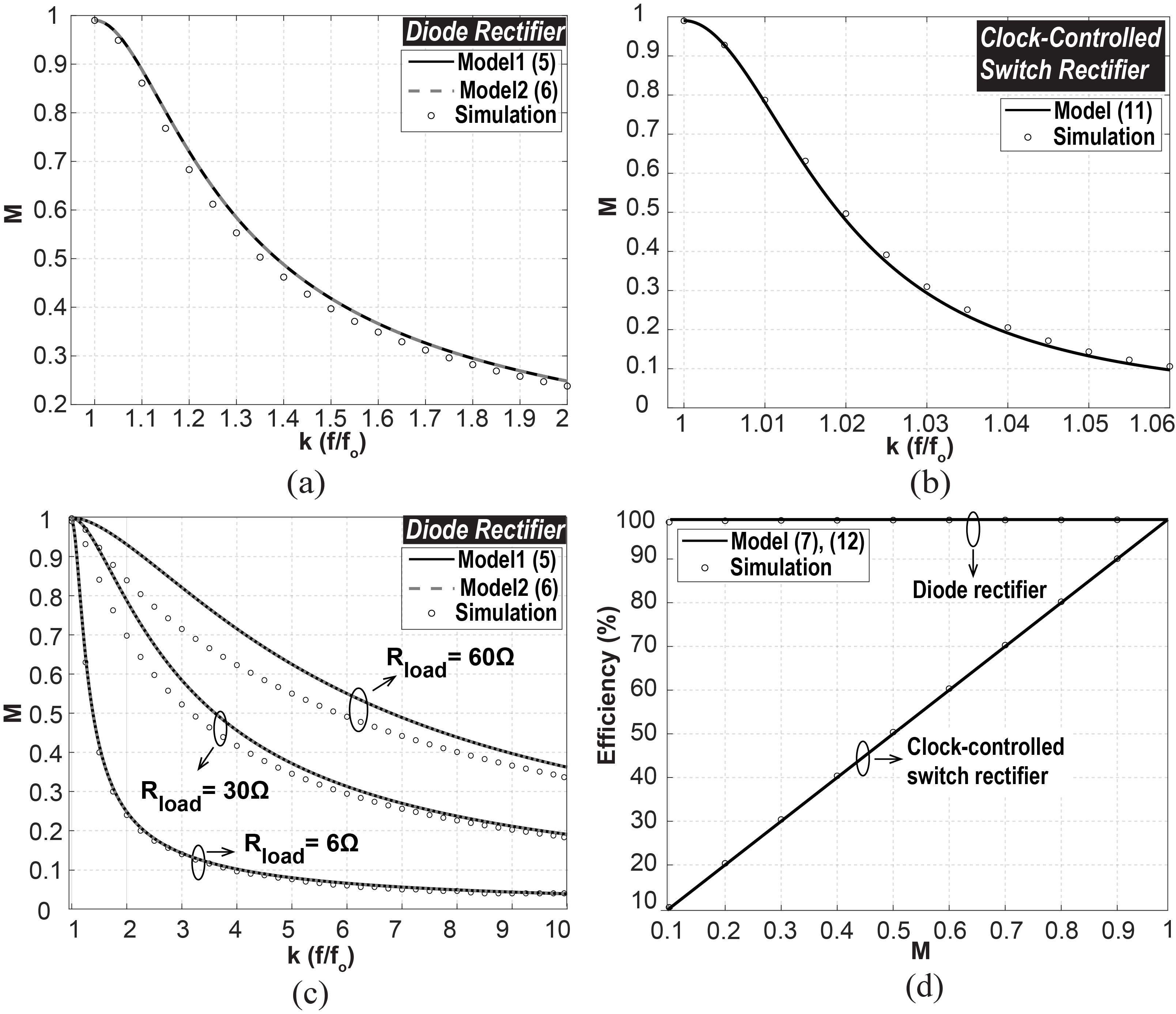}
\caption{Comparison between model predictions and simulation results for the conversion ratio, $M$, versus the normalized switching frequency, $k = f/f_o$, for $V_{in}$ = 12\,V and $f_o$ = 503\,kHz: (a) Conventional series-resonant converter~\eqref{eq:Mconvdiode_secondmethod} and~\eqref{eq:M1x1resc}, (b) series-resonant converter with a clock-controlled switch rectifier~\eqref{eq:Mwodiodes}, and (c) series-resonant converter under different values of $R_{load}$. (d) Efficiency versus conversion ratio for a conventional series-resonant converter employing an uncontrolled~\eqref{eq:etaconvseries} and a controlled rectifier~\eqref{eq:eta1xvseries_withoutdiodes}.}
\label{fig.1x1model_vs_sim}
\end{figure}
\begin{figure}[!tb]
\centering
\includegraphics[width=2.4in]{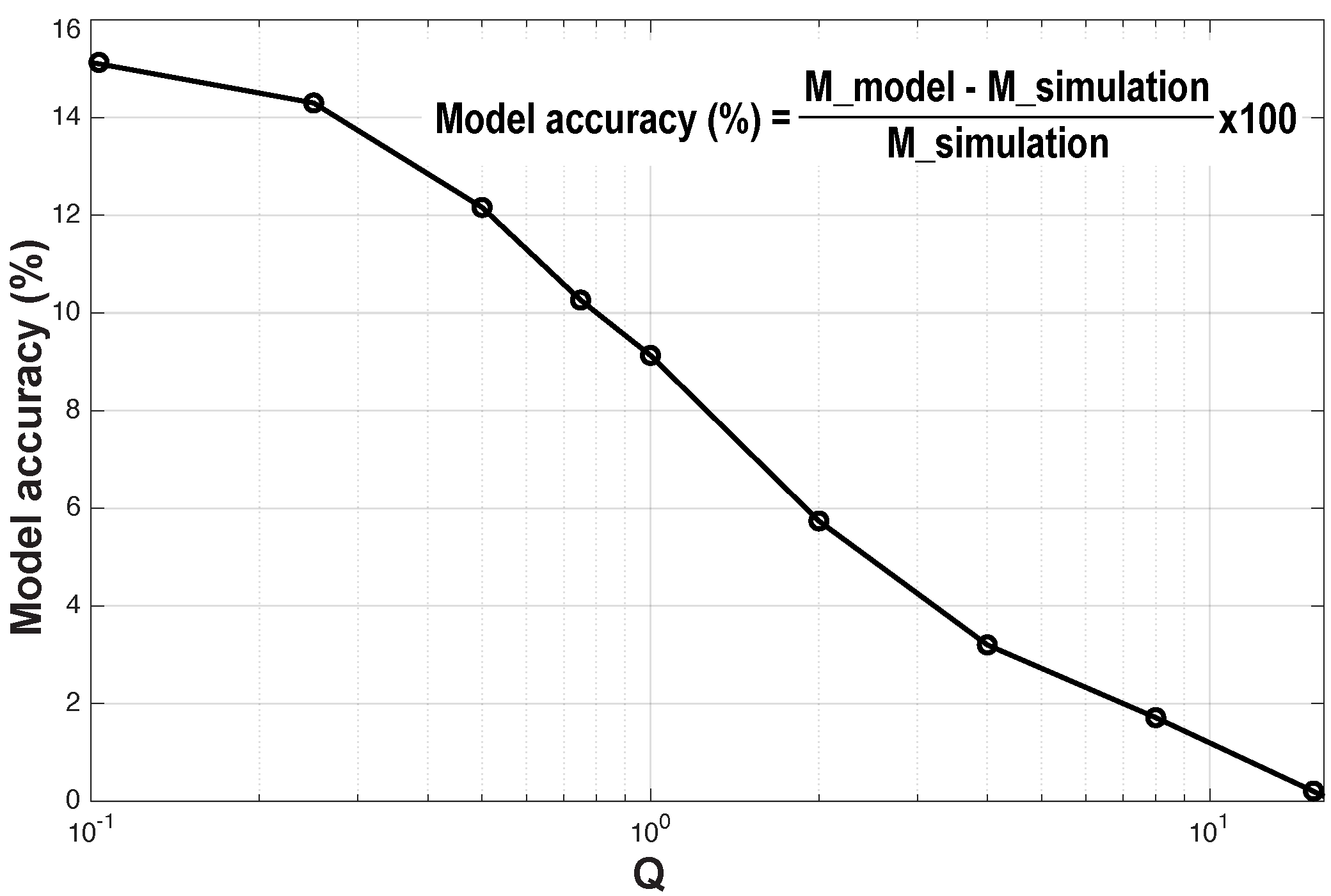}
\caption{\textcolor{black}{Model prediction accuracy of the conversion ratio versus quality factor Q for the series-resonant converter at full load.} }\label{fig.qerror}
\end{figure}

Fig.~\ref{fig.1x1model_vs_sim}(a) and (b) compare the conversion ratios of a series-resonant converter using a diode rectifier and a controlled rectifier, as given by the analytical expressions in~\eqref{eq:Mconvdiode_secondmethod},~\eqref{eq:M1x1resc}, and~\eqref{eq:Mwodiodes}, with simulation results under $R_{load}$\,=\,6\,$\Omega$. The analytical predictions are within 3.7$\%$ from the simulation results for $r_{ON}$= 6\,m$\Omega$ and $Q$\,=\,2.6, where $Q$ is defined as \(\frac{1}{R_e}\sqrt{\frac{L}{C}}\). The model deviates from simulation results as $R_{load}$ is increased to 30\,$\Omega$ and 60\,$\Omega$ due to the lower $Q$ values of 0.52 and 0.26, respectively, as shown in Fig.~\ref{fig.1x1model_vs_sim}(c). At low $Q$ values, the tank response differs from the assumed purely sinusoidal waveform, where higher-order harmonics start to appear in the tank response. Fig.~\ref{fig.1x1model_vs_sim}(d) compares the analytical expressions in \eqref{eq:etaconvseries} and \eqref{eq:eta1xvseries_withoutdiodes} against simulation results under $R_{load}$\,=\,6\,$\Omega$. \textcolor{black}{Fig. \ref{fig.qerror} shows the model prediction accuracy of the conversion ratio versus the quality factor for the series-resonant converter at full load. The model deviates by up to 15.1\% at very low quality factors, while the deviation decreases to 0.1\% at high quality factors.}

\begin{figure}[!tb]
\centering
\includegraphics[width=2.8in]{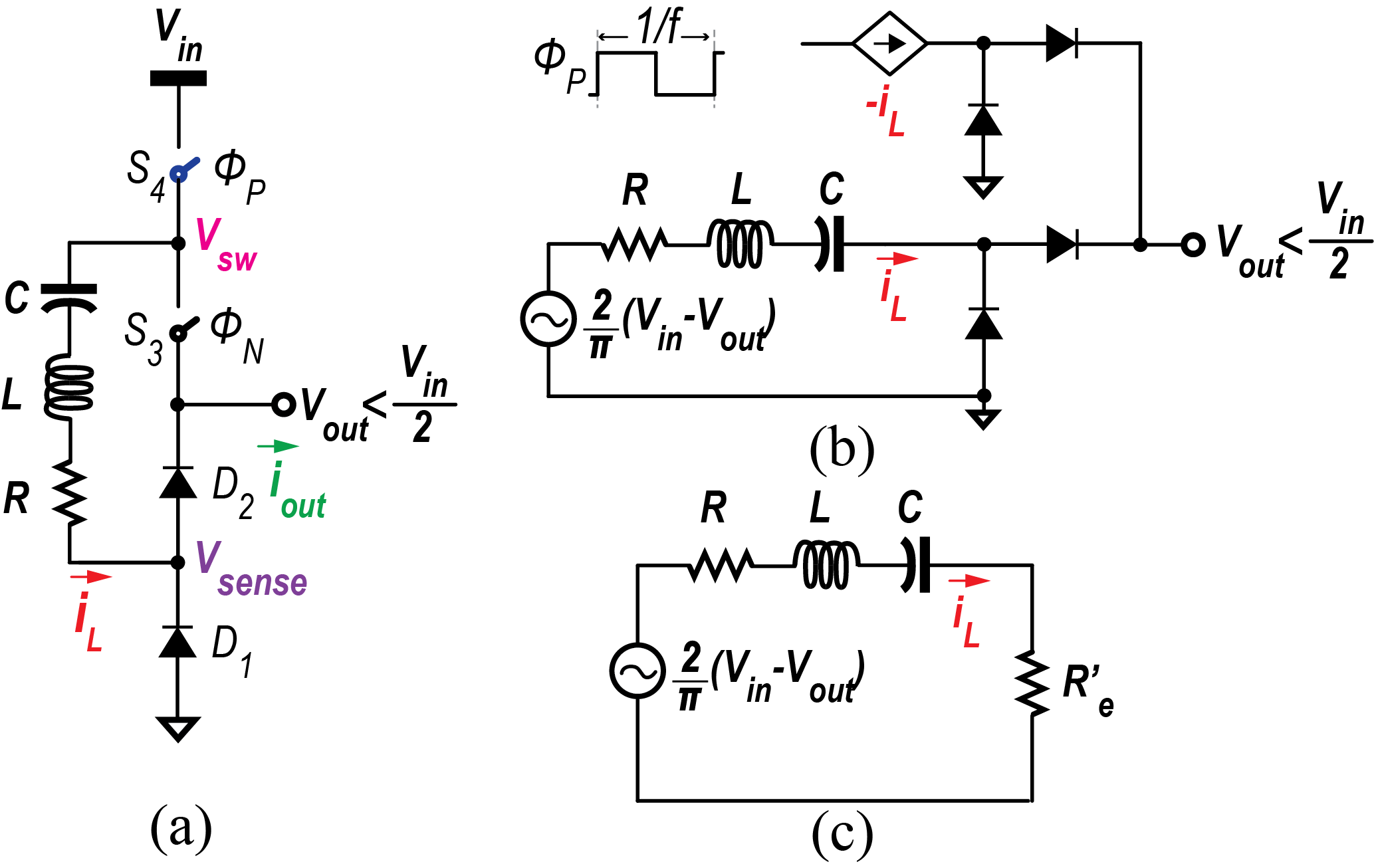}
\caption{(a) The switch diagram of a 2:1 ReSC converter. (b) Equivalent circuit representation illustrating full-wave rectification. (c) Steady-state equivalent circuit of a 2:1 resonant SC converter that models the behavior of the converter around the fundamental component of $V_{sw}$.}
\label{fig.2x1ReSC}
\end{figure}
\subsection{Loss Analysis of a 2:1 ReSC dc-dc Converter}
\begin{figure}[!tb]
\centering
\includegraphics[width=3.4in]{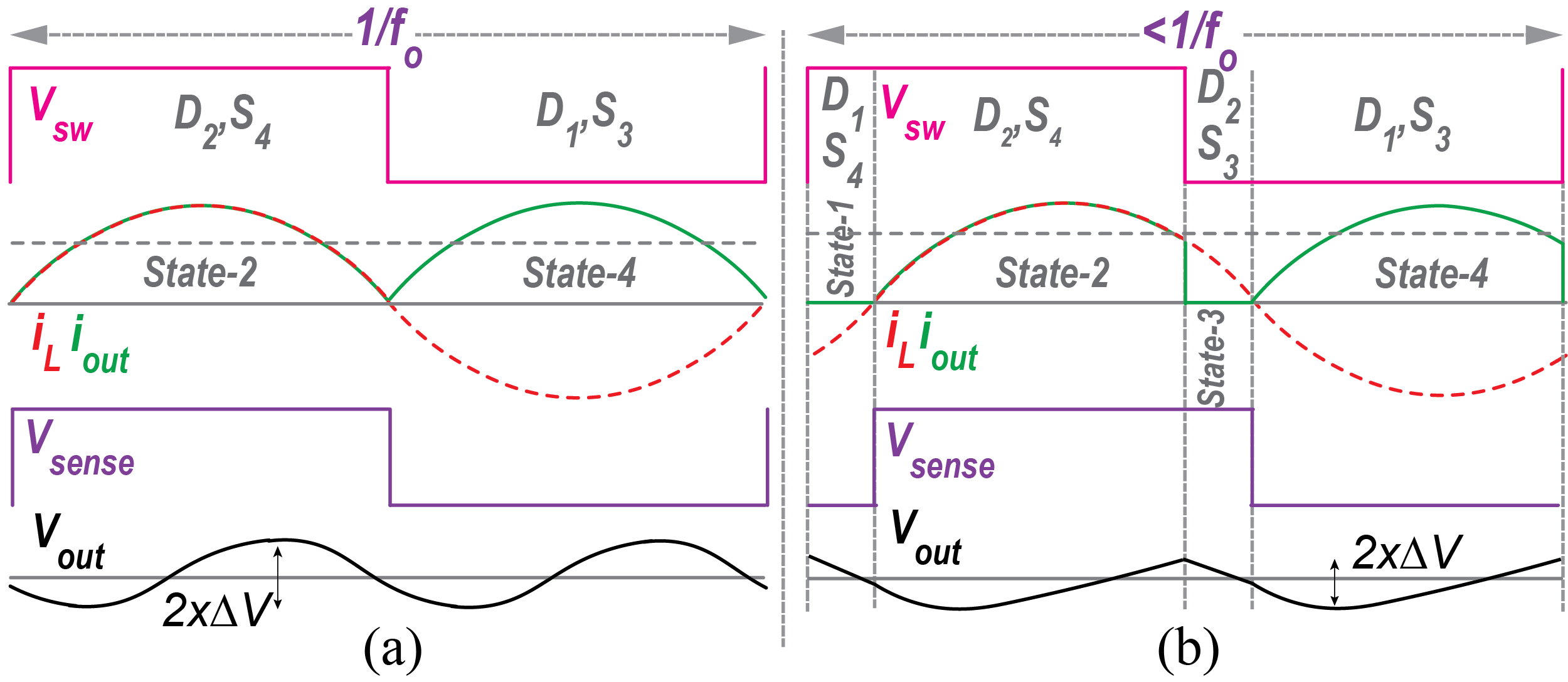}
\caption{2:1 resonant SC converter with diode-operated bottom switches simulated switching-node voltage, $V_{sw}$, tank current, $i_L$, output current, $i_{out}$, and output voltage, $V_{out}$, (a) at resonance and (b) above resonance, $f > f_{o}$.}
\label{fig.2x1ReSC_timingdiagram}
\end{figure}

Fig.~\ref{fig.2x1ReSC}(a) shows the schematic of a 2:1 resonant SC converter. At resonance, $f$\,=\,$f_{o}$, the tank current in a 2:1 ReSC converter is delivered to the output across the entire clock cycle, which transforms the bottom half-bridge rectifier into a full-wave rectifier, as illustrated in Fig.~\ref{fig.2x1ReSC}(b) and Fig.~\ref{fig.2x1ReSC_timingdiagram}(a).
By operating the bottom MOSFETs, $S_{1,2}$, as diodes, $D_{1,2}$, the switching node voltage of the bottom half bridge in a 2:1 SC resonant converter, $V_{sense}$, is kept in phase with the inductor current, $i_L$, under different $f$ values, which eliminates the imaginary part of $Z_{e}$, as shown in Fig.~\ref{fig.2x1ReSC}(c), and minimizes the converter conduction loss. 
In steady state and under a high $Q$ factor, the nearly sinusoidal tank current can be expressed as
{%
\setlength{\abovedisplayskip}{0.5\baselineskip}%
\setlength{\belowdisplayskip}{0.5\baselineskip}%
\begin{equation}
\begin{aligned}
i_{L}(t) &= \frac{2}{\pi |Z_o|} \left[
(V_{in}-V_{out}) \sin\left(\omega t - \theta \right) \right. \\
&\quad \left. - V_{out} \sin\left(\omega t - \phi - \theta \right)
\right].
\end{aligned}
\label{eq:iL2x1secondmethod}
\end{equation}
}Similar to \eqref{eq:1xconveq}, the tank current of a 2:1 ReSC also can be expressed as 
{%
\setlength{\abovedisplayskip}{0.5\baselineskip}%
\setlength{\belowdisplayskip}{0.5\baselineskip}%
\begin{equation}
\begin{aligned}
i_{L}(t) &= \frac{2(V_{in}-V_{out})}{\pi |Z|}\sin\left(\omega t - \phi \right),
\label{eq:2xconveq} 
\end{aligned}
\end{equation}}where $|Z|$ is the magnitude of the total impedance, $Z$, presented to the top half-bridge inverter by the resonant tank, $Z_o$, and the bottom rectifier, $Z^{'}_e$, i.e., $Z = Z_o + Z^{'}_e$. The phase of $Z$, denoted by $\phi$, represents the phase lag between the tank current and the top half-bridge switching clock. Note, $Z$ is given by
{\setlength{\abovedisplayskip}{0.5\baselineskip}%
\setlength{\belowdisplayskip}{0.5\baselineskip}%
\begin{equation}
\begin{aligned}
|Z| &= \sqrt{\left(R+R^{'}_{e}\right)^2 + \left(\omega L - \frac{1}{\omega C}+X^{'}_e\right)^2},\\
\phi &= \tan^{-1} \left( \frac{\omega L - \frac{1}{\omega C} + X^{'}_e}{R+R^{'}_{e}} \right).
\label{eq:2xconvimpeq} 
\end{aligned}
\end{equation}}

The dc output voltage is given by
{%
\setlength{\abovedisplayskip}{0.5\baselineskip}%
\setlength{\belowdisplayskip}{0.5\baselineskip}%
\begin {equation}
V_{out}=\frac{|i_{L}|R_{load}}{\pi}(1+\cos(\phi)).
\label{eq:vout2x1resc}
\end {equation}}Note that, $V_{out}$ is a function of $\phi$ unlike the conventional case, since for $f$\,$>$\,$f_{o}$ two additional states, state-1 and state-3, that depend on $\phi$ are introduced, as shown in Fig.~\ref{fig.2x1ReSC_timingdiagram}(b). These states apply a short across the tank or connect it across $V_{in}$, and hence, lower the amount of charge delivered to the output per clock cycle as well as the $V_{out}$ value.
From~\eqref{eq:iL2x1secondmethod} and~\eqref{eq:vout2x1resc}, the converter dc conversion ratio, $M$, can be found as 
{%
\setlength{\abovedisplayskip}{0.5\baselineskip}%
\setlength{\belowdisplayskip}{0.5\baselineskip}%
\begin{equation}
\begin{aligned}
M &= \left[1+
\frac{1}{\sqrt{1 + T^2}} \left(
1 + \frac{T \left(1 + \tan\theta \times T\right)}{ \tan\theta - T }
\right)
\right]^{-1}, \\
T &= \frac{ \sin\theta }{ \cos\theta + \dfrac{2 R_{load}}{\pi^2 |Z_o|} \left(1 + \dfrac{1}{\sqrt{1 + T^2}} \right) }.
\end{aligned}
\label{eq:M2x1diode_secondmethod}
\end{equation}}
The bottom diodes shown in Fig.~\ref{fig.2x1ReSC}(a) eliminate the imaginary part, $X^{'}_e$, of $Z^{'}_e$. The 2:1 ReSC can then be modeled by the equivalent circuit shown in Fig.~\ref{fig.2x1ReSC}(c). The $R{'}_{e}(\omega)$ value in the case of a 2:1 ReSC is that of a conventional series-resonant converter but multiplied by \((1+\cos(\phi))\), i.e., \(\frac{2}{\pi^2}R_{load}(1+\cos(\phi))\). $R{'}_{e}(\omega)$ is a function of $\phi$, and hence, $f$, due to the additional no-output-charge state-1 and state-3 in Fig.~\ref{fig.2x1ReSC_timingdiagram}(b). From~\eqref{eq:2xconveq} and~\eqref{eq:vout2x1resc}, $M$ can be found as
{%
\setlength{\abovedisplayskip}{0.5\baselineskip}%
\setlength{\belowdisplayskip}{0.5\baselineskip}%
\begin {equation}
M = \left(1+\frac{|Z|}{R^{'}_{e}(\omega)}\right)^{-1},
\label{eq:M2x1resc}
\end {equation}}which has a similar form to the conversion ratio of a conventional series-resonant converter in ~\eqref{eq:M1x1resc}, but with an additional 1 in the denominator to provide an $M$\,$\sim$\,1/2 at resonance. 
The 2:1 ReSC efficiency is given by 
{%
\setlength{\abovedisplayskip}{0.5\baselineskip}%
\setlength{\belowdisplayskip}{0.5\baselineskip}%
\begin {equation}
\eta = \left(1 + \frac{R}{R_{e}(\omega)}\right)^{-1},
\label{eq:eta2x1ReSC}
\end {equation}}where $R_{e}(\omega)$ represents the ac-side effective resistance of $R_{load}$ resulting from the application of ac-to-dc power balance, i.e., \( \frac{1}{2} ~ |i_L|^2 R_e(\omega) = V_{out}^2/R_{load}\), and hence, $R_{e}(\omega)$ is equal to $R^{'}_{e}(\omega)(1+\cos(\phi)$. $R^{'}_e$ only considers the tank current delivered to $R_{load}$ within a half-cycle time rather than the entire switching cycle as $R_e$.   
Unlike the conventional series-resonant converter, the efficiency of a 2:1 ReSC is not constant versus M under a given $R_{load}$ and $R$, but rather a weak function of the phase lag, $\phi$.
\begin{figure}[!tb]
  \centering
  \includegraphics[width=2.3in]{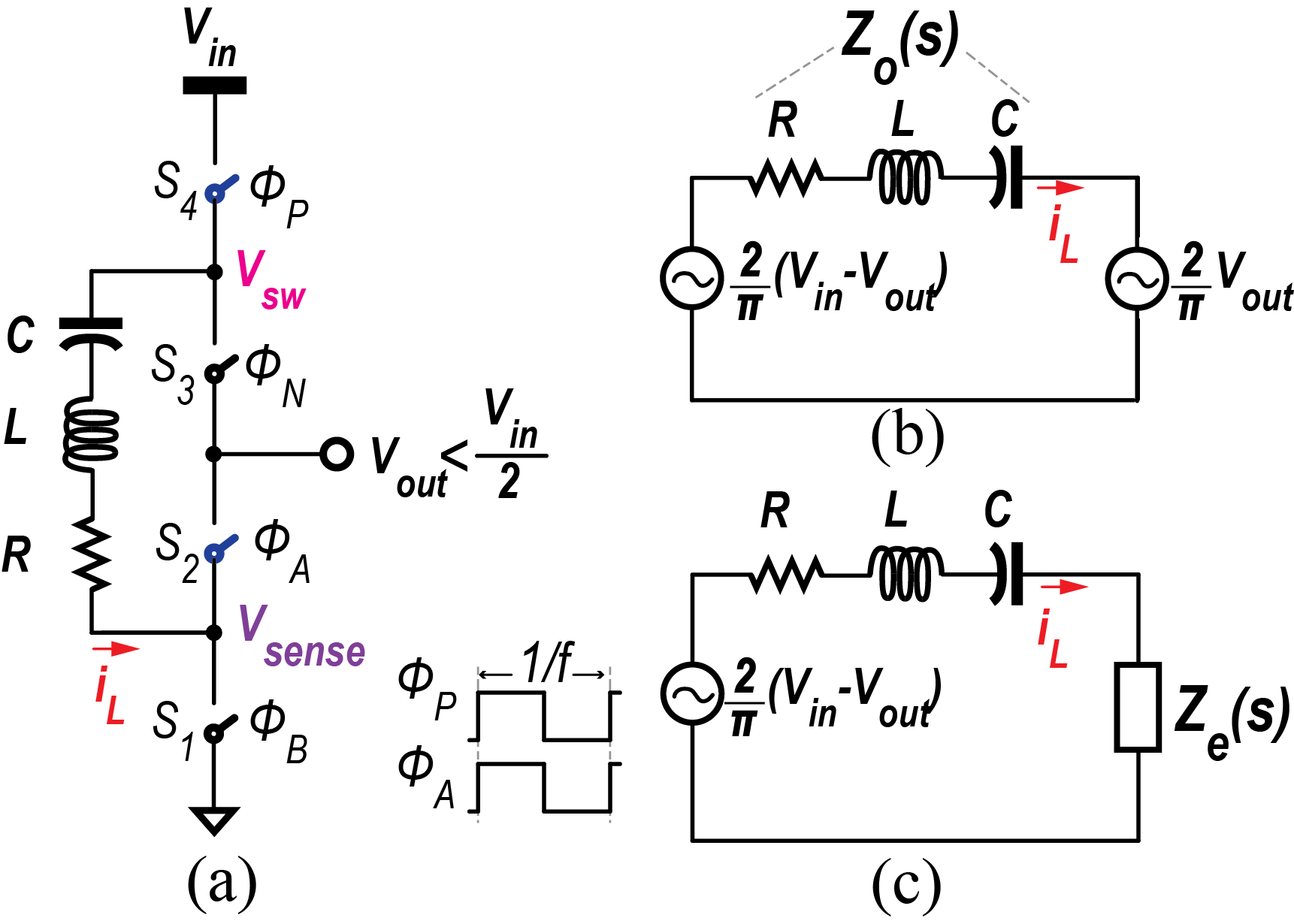}
  \caption{(a) The schematic diagram of a 2:1 resonant SC converter with a clock-controlled switch rectifier. (b) Steady-state equivalent circuit near resonance. (c) Steady-state equivalent circuit of the converter with the rectifier modeled as a reactive impedance.}
 \label{fig.2x1ReSC_without_diodes}
\end{figure}

As shown in Fig.~\ref{fig.2x1ReSC_without_diodes}, prior 2:1 ReSC converters typically operate the bottom two switches, $S_{1,2}$, from a clock that is in phase with the top half bridge, $S_{3,4}$, clock rather than as diodes. In this case, the rectifier in a 2:1 ReSC converter is operated as the controlled rectifier discussed in section~\ref{subsec:conventional_ReSC_loss_analysis} rather than as the suggested uncontrolled rectifier, and hence, the 2:1 ReSC rectifier in this case presents a reactive impedance $Z_{e}(s)$ given by   
{%
\setlength{\abovedisplayskip}{0.5\baselineskip}%
\setlength{\belowdisplayskip}{0.5\baselineskip}%
\begin{equation}
\begin{aligned}
  Z_{e}(s) &= \frac{8}{\pi^2}R_{load} \cos(\phi) \phase{\phi}\\
           &= \frac{M}{1 - 2M} Z_o(s) ~ \text{for} ~ M < \frac{1}{2}.
\end{aligned}
\label{eq:zwodiodes2x1}
\end{equation}}

The tank current can be approximated by a sinusoidal waveform as in~\eqref{eq:iL2x1secondmethod} with $\phi$ in the second term set to 0, according to the equivalent circuit in Fig.~\ref{fig.2x1ReSC_without_diodes}(b), or~\eqref{eq:2xconveq} according to the model in Fig.~\ref{fig.2x1ReSC_without_diodes}(c).
Similar to~\eqref{eq:zwodiodes}, as the phase lag, $\phi$, increases with higher $f$, the magnitude of $Z_e$ in~\eqref{eq:zwodiodes2x1} decreases, which increases the tank current for the same $R_{load}$ and leads to greater conduction loss.
The dc output voltage is given by
{%
\setlength{\abovedisplayskip}{0.5\baselineskip}%
\setlength{\belowdisplayskip}{0.5\baselineskip}%
\begin{equation} \begin{aligned} 
&V_{out}=\frac{2|i_{L}|R_{load}}{\pi}\cos(\phi), 
\end{aligned}
\label{eq:voutrescwodiodes1}
\end{equation}}which is twice the value of~\eqref{eq:phiwodiodes} due to full-wave rectification. 
The dc conversion ratio, $M$, of the converter can be found from~\eqref{eq:voutrescwodiodes1} along with~\eqref{eq:iL2x1secondmethod} or~\eqref{eq:2xconveq}, as 
{%
\setlength{\abovedisplayskip}{0.5\baselineskip}%
\setlength{\belowdisplayskip}{0.5\baselineskip}%
\begin{equation}
M = \frac{1}{2}\left(1+\frac{R}{R_e(\omega)}\right)^{-1}.
\label{eq:M2x1rescwodiodes}
\end{equation}}
The efficiency of the converter is given by
{%
\setlength{\abovedisplayskip}{0.5\baselineskip}%
\setlength{\belowdisplayskip}{0.5\baselineskip}%
\begin{equation} \eta = \left(1 + \frac{R}{R_e(\omega)} \right)^{-1} = 2M, \label{eq:eta2x1ReSCwithoutdiodes} \end{equation}}where $R_e(\omega)$ is the real part of $Z_e$ which is equal to ~$\frac{8}{\pi^2} R_{load}\cos^2(\phi)$. 

\begin{figure}[!tb]
\centering
\includegraphics[width=3.4in]{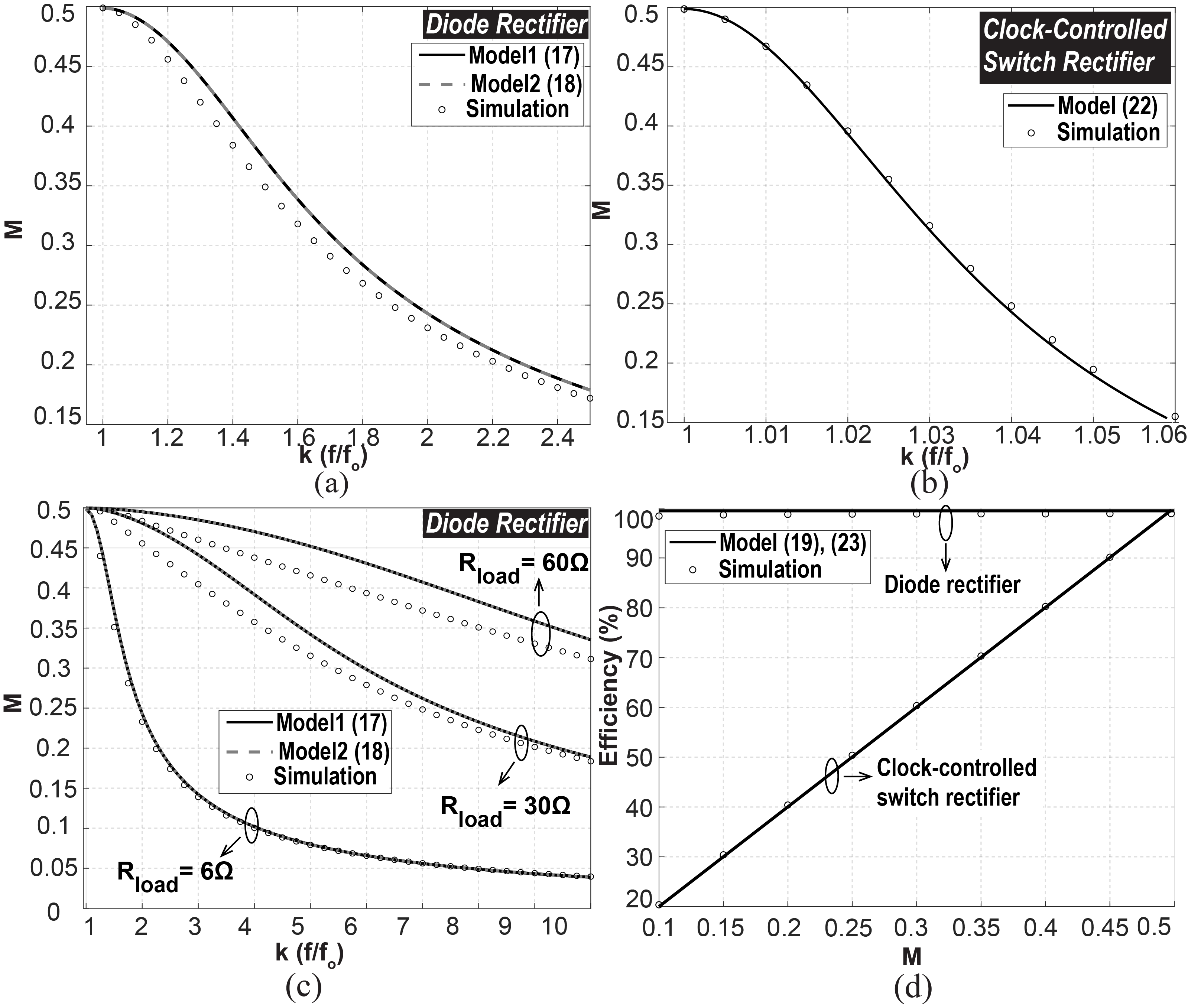}
\caption{Comparison between model predictions and simulation results for the conversion ratio, $M$, versus the normalized switching frequency, $k = f/f_o$, for $V_{in}$ = 24\,V and $f_o$ = 503\,kHz: (a) 2:1 ReSC converter with diode-operated bottom switches\,\eqref{eq:M2x1diode_secondmethod} and\,\eqref{eq:M2x1resc}, (b) 2:1 ReSC converter with a clock-controlled switch rectifier\,\eqref{eq:M2x1rescwodiodes}, and (c) 2:1 ReSC converter under different values of $R_{load}$. (d) Efficiency versus M for a 2:1 ReSC converter with diode-operated bottom switches\,\eqref{eq:eta2x1ReSC} and a clock-controlled bottom switches\,\eqref{eq:eta2x1ReSCwithoutdiodes}.}
\label{fig.2x1model_vs_sim}
\end{figure}

Fig.~\ref{fig.2x1model_vs_sim}(a) and (b) compare the conversion ratios of a 2:1 ReSC converter with a diode rectifier and a controlled rectifier, the analytical expressions in~\eqref{eq:M2x1diode_secondmethod},~\eqref{eq:M2x1resc}, and~\eqref{eq:M2x1rescwodiodes}, to simulation results under $R_{load}$\,=\,6\,$\Omega$. The analytical models are within 1$\%$ of the simulation results for $r_{ON}$\,=\,6\,m$\Omega$ and $Q$\,=\,1.3, where $Q$ is defined as \( \frac{1}{R'_e} \sqrt{\frac{L}{C}} \), and $R'_e$ is the value of $R'_e(\omega)$ evaluated at $f_o$. Fig.~\ref{fig.2x1model_vs_sim}(c) shows the conversion ratio of a 2:1 ReSC converter under $R_{load}$\,=\,6\,$\Omega$, 30\,$\Omega$, and 60\,$\Omega$. Fig.~\ref{fig.2x1model_vs_sim}(d) compares the analytical expressions in \eqref{eq:eta2x1ReSC} and \eqref{eq:eta2x1ReSCwithoutdiodes} to simulation results under $R_{load}$\,=\,6\,$\Omega$. 

\begin{figure}[!tb]
\centering
\includegraphics[width=2.8in]{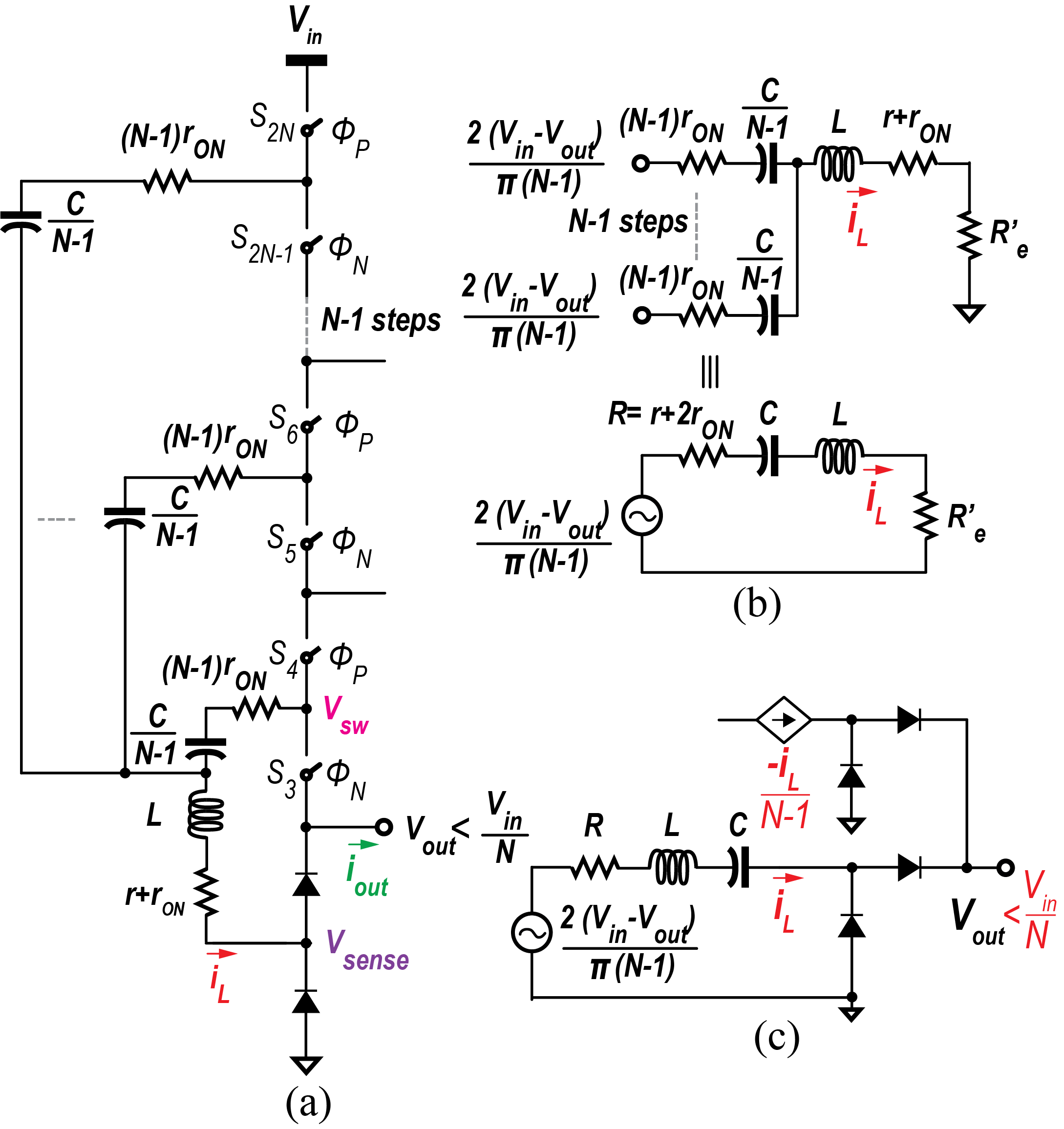}
\caption{(a) The switch diagram of an $N$:1 ReSC converter. The fixed ladder is not shown for simplicity. (b) Steady-state equivalent circuit near resonance. (c) Equivalent circuit representation illustrating full-wave rectification.}
\label{fig.Nx1ReSC}
\end{figure}

\subsection{Loss Analysis of an $N$:1 ReSC Ladder Converter}

\begin{figure}[!tb]
\centering
\includegraphics[width=3.4in]{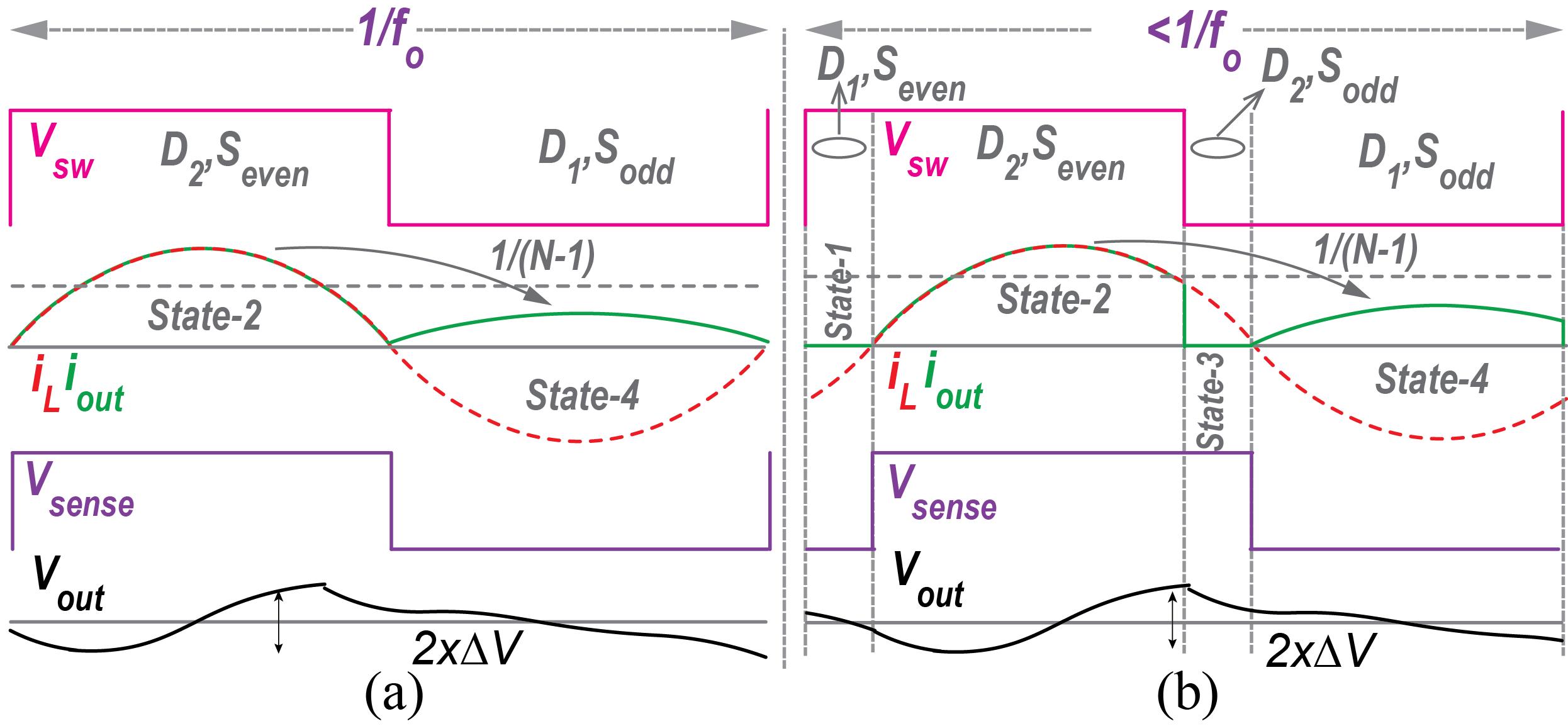}
\caption{$N$:1 ReSC converter simulated switching-node voltage, $V_{sw}$, tank current, $i_L$, output current, $i_{out}$, and output voltage, $V_{out}$, (a) at resonance, $f = f_{o}$, and (b) above resonance, $f > f_{o}$.}
\label{fig.Nx1ReSC_timingdiagram}
\end{figure}

~~The preceding analysis can be extended to an $N$:1 ReSC ladder converter with a diode rectifier and a clock-controlled rectifier in the bottom half-bridge, as shown in Fig.~\ref{fig.Nx1ReSC}(a). The analysis in this section is only valid for $N\geq2$. 
The switch on-resistance, $(N-1)r_{ON}$, of the upper $N-1$ half bridges and their flying capacitors, $C/(N-1)$ can be represented by a single resistance, $r_{ON}$, and capacitance, $C$, as depicted in Fig.~\ref{fig.Nx1ReSC}(b). The resistor $r+r_{ON}$ on the output side in Fig.~\ref{fig.Nx1ReSC}(a) and (b) represents the ESR of the inductor, L, and flying capacitor, C, in addition to the equivalent diode-operated switches resistance, assuming that each diode-operated switch has an on-resistance of $r_{ON}$. Hence, the total lumped resistance in series with the LC tank, $R$, is equal to $r+2r_{ON}$. The steady-state waveforms of the converter at the resonant frequency, $f_o$, and above the resonant frequency, $f>f_o$, are illustrated in Fig.~\ref{fig.Nx1ReSC_timingdiagram}(a) and (b), respectively. The four states of operation of the circuit are similar to a 2:1 ReSC converter, with the difference that state-4 delivers $N-1$ lower charge to the output load, compared to state-2. 
In steady state and under a high $Q$ factor, the nearly sinusoidal tank current can be expressed as
{%
\setlength{\abovedisplayskip}{0.5\baselineskip}%
\setlength{\belowdisplayskip}{0.5\baselineskip}%
\begin{equation}
\begin{aligned}
i_{L}(t) &= \frac{2}{\pi |Z_o|} [
\frac{(V_{in}-V_{out})}{(N-1)} \sin\left(\omega t - \theta \right) \\
&\quad - V_{out} \sin\left(\omega t - \phi - \theta \right)
],
\end{aligned}
\label{eq:iLNx1secondmethod}
\end{equation}}where $Z_o$, $\theta$, and $\phi$ are as defined following~\eqref{eq:secondmethod} in \ref{subsec:conventional_ReSC_loss_analysis}.

Analogous to~\eqref{eq:1xconveq} and~\eqref{eq:2xconveq}, the tank current of an $N$:1 ReSC converter is given by
{%
\setlength{\abovedisplayskip}{0.5\baselineskip}%
\setlength{\belowdisplayskip}{0.5\baselineskip}%
\begin{equation}
\begin{aligned}
i_{L}(t) &= \frac{2(V_{in}-V_{out})}{\pi |Z|(N-1)}\sin\left(\omega t - \phi \right),
\label{eq:Nxconveq} 
\end{aligned}
\end{equation}}where $|Z|$ is the magnitude of the total impedance, $Z$, seen by the top $N$-1 half bridges in parallel, as shown in Fig.~\ref{fig.Nx1ReSC}(b), which is equal to $Z_o$ in addition to the bottom rectifier equivalent $Z^{'}_e$, as given by~\eqref{eq:2xconvimpeq}. Similar to a 2:1 ReSC analysis, both relations~\eqref{eq:iLNx1secondmethod} and~\eqref{eq:Nxconveq} for the tank current are used to find two expressions for the conversion ratio, $M$, of the $N$:1 resonant SC converter shown in Fig.~\ref{fig.Nx1ReSC}(a). As will be demonstrated, both expressions yield identical results.
For an $N$:1 converter, the dc output voltage is given by 
{%
\setlength{\abovedisplayskip}{0.5\baselineskip}%
\setlength{\belowdisplayskip}{0.5\baselineskip}%
\begin{equation}
V_{out}=\frac{N}{2(N-1)}\left[ \frac{|i_L|R_{load}}{\pi}(1+\cos(\phi))\right].
\label{eq:voutNx1resc}
\end{equation}}Under a higher $V_{in}$ that maintains the voltage swing of the half-bridge inverters in an $N$:1 ReSC identical to that of a 2:1 ReSC, as $N$ is increased the charge delivered during state-4 decreases, and in the limit, the $N$:1 ReSC converter provides half of the charge compared to a 2:1 ReSC, according to~\eqref{eq:voutNx1resc}. 
From~\eqref{eq:iLNx1secondmethod} and~\eqref{eq:voutNx1resc}, the converter dc conversion ratio, $M$, can be found as 
{%
\setlength{\abovedisplayskip}{0.5\baselineskip}%
\setlength{\belowdisplayskip}{0.5\baselineskip}%
\begin{equation}
\begin{aligned}
M &= \left( 1 +  \frac{(N - 1)}{\sqrt{1 + T^2}}  \left( 1 + \frac{T (1 + T \times\tan\theta)}{\tan\theta - T} \right) \right)^{-1}, \\
T &= \frac{ \sin\theta }{ \cos\theta + \left( \frac{R_{load}}{\pi^2 |Z_o|} \left(1 + \dfrac{1}{\sqrt{1 + T^2}} \right) \frac{N}{(N - 1)} \right) }.
\end{aligned}
\label{eq:MNx1diode_secondmethod}
\end{equation}}

The $N$:1 ReSC converter shown in Fig.~\ref{fig.Nx1ReSC}(a) can be modeled by the equivalent circuit shown in Fig.~\ref{fig.Nx1ReSC}(b), where the uncontrolled bottom half-wave rectifier always presents a real load, $R^{'}_e(\omega)$, to the tank that is equal to \(\frac{N}{(N-1)}\frac{R_{load}}{\pi^2}(1+\cos(\phi))\).
In this case, from~\eqref{eq:Nxconveq} and~\eqref{eq:voutNx1resc}, $M$ can be found as
{%
\setlength{\abovedisplayskip}{0.5\baselineskip}%
\setlength{\belowdisplayskip}{0.5\baselineskip}%
\begin{equation}
M = \left(1+(N-1)\frac{|Z|}{R^{'}_{e}(\omega)}\right)^{-1}.
\label{eq:MNx1resc}
\end{equation}}From ac-to-dc power balance, the $R_{e}(\omega)$ value in the case of an $N$:1 ReSC can be found as 
{%
\setlength{\abovedisplayskip}{0.5\baselineskip}%
\setlength{\belowdisplayskip}{0.5\baselineskip}%
\begin{equation}
\begin{aligned}
R_{e}(\omega) &= \frac{N^{2}}{2(N-1)^{2}}\frac{R_{load}}{\pi^2}(1+\cos(\phi))^{2}.
\end{aligned}
\end{equation}}
The $N$:1 ReSC efficiency is given by
{%
\setlength{\abovedisplayskip}{0.5\baselineskip}%
\setlength{\belowdisplayskip}{0.5\baselineskip}%
\begin{equation}
\eta = \left(1 + \frac{R}{R_{e}(\omega)}\right)^{-1}.
\label{eq:etaNx1ReSC}
\end{equation}}

Analogously, the model of a 2:1 ReSC converter with a controlled rectifier can be extended to an $N$:1 converter.
The impedance $Z_e(s)$ is given by
{%
\setlength{\abovedisplayskip}{0.5\baselineskip}%
\setlength{\belowdisplayskip}{0.5\baselineskip}%
\begin{equation} \begin{aligned} 
  Z_e(s) &= \frac{2N^{2}}{(N - 1)^{2} \pi^2} R_{load} \cos(\phi) \phase{\phi}\\
  &= \frac{(N-1)M}{1-NM}Z_o(s)  ~ \text{for} ~ M< \frac{1}{N}.
  \label{eq:zwodiodesNx1}
\end{aligned}\end{equation}}Similar to a 2:1 ReSC with a controlled rectifier, the tank current of an $N$:1 converter can be approximated by a sinusoidal waveform, as expressed in~\eqref{eq:iLNx1secondmethod}, with $\phi = 0$. The dc output voltage is given by
{%
\setlength{\abovedisplayskip}{0.5\baselineskip}%
\setlength{\belowdisplayskip}{0.5\baselineskip}%
\begin{equation} 
\begin{aligned}
&V_{out}=\frac{N}{(N-1)}\frac{|i_{L}|R_{load}}{\pi}\cos(\phi).
\label{eq:voutNx1rescwodiodes}
\end{aligned} 
\end{equation}}$M$ of the converter can be found from~\eqref{eq:voutNx1rescwodiodes} along with~\eqref{eq:iLNx1secondmethod} or~\eqref{eq:Nxconveq}, as 
{%
\setlength{\abovedisplayskip}{0.5\baselineskip}%
\setlength{\belowdisplayskip}{0.5\baselineskip}%
\begin{equation} 
\begin{aligned}
M = \frac{1}{N}\left(1+\frac{R}{R_e(\omega)}\right)^{-1},
\label{eq:MNx1rescwithoutdiodes}
\end{aligned} 
\end{equation}}where $R_e(\omega)$ is the real part of $Z_e$.
The efficiency of the converter is given by
{%
\setlength{\abovedisplayskip}{0.5\baselineskip}%
\setlength{\belowdisplayskip}{0.5\baselineskip}%
\begin{equation} 
\begin{aligned}
&\eta = \left(1 + \frac{R}{R_{e}(\omega)}\right)^{-1} = NM.
\label{eq:etaNx1ReSCwodiodes}
\end{aligned} 
\end{equation}}

\begin{figure}[!tb]
\centering
\includegraphics[width=3.4in]{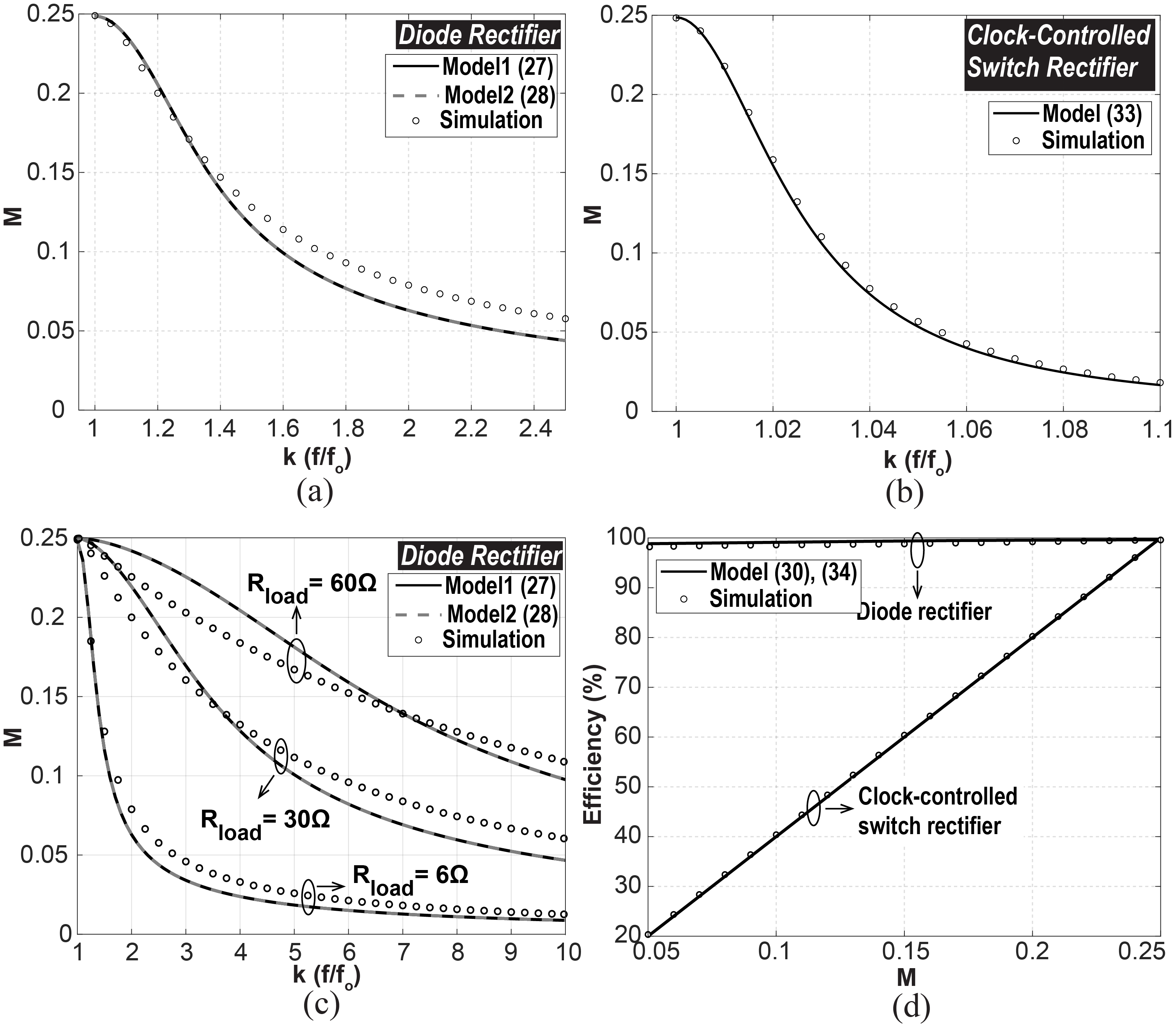}
\caption{Comparison of model and simulation results for the conversion ratio, $M$, versus normalized switching frequency, $k = f/f_o$, for $V_{in}$ = 48\,V and $f_o$ = 503\,kHz: (a) 4:1 resonant SC converter \eqref{eq:MNx1diode_secondmethod} and \,\eqref{eq:MNx1resc}, (b) 4:1 resonant SC converter with a clock-controlled switch rectifier\,\eqref{eq:MNx1rescwithoutdiodes}, and (c) 4:1 resonant SC converter for different values of $R_{load}$. (d) Efficiency versus conversion ratio for the 4:1 resonant SC converter with a diode rectifier\,\eqref{eq:etaNx1ReSC} and with a clock-controlled switch rectifier\,\eqref{eq:etaNx1ReSCwodiodes}.}\label{fig.4x1model_vs_sim}
\end{figure}

Fig.~\ref{fig.4x1model_vs_sim}(a), (b), and (d) illustrate the comparison between the analytical expressions in~\eqref{eq:MNx1diode_secondmethod},~\eqref{eq:MNx1resc},~\eqref{eq:etaNx1ReSC},~\eqref{eq:MNx1rescwithoutdiodes}, and~\eqref{eq:etaNx1ReSCwodiodes} and the corresponding simulation results for $R_{load}$\,=\,6\,$\Omega$. 
The analytical predictions are within 1.68$\%$ from the simulation results for $r_{ON}$\,=\,6\,m$\Omega$ and $Q$\,=\,1.95. 
Fig.~\ref{fig.4x1model_vs_sim}(c) presents the results for $M$ under $R_{load}$\,=\,6\,$\Omega$, 30\,$\Omega$, and 60\,$\Omega$. 
Table \ref{tab:tbw} summarizes the analytical expressions found in this section.

\begin{table*}[t]
    \centering
    \caption{Steady-State Expressions for Output Voltage, Conversion Ratio, and Efficiency of ReSC Converters With uncontrolled and Controlled Rectifiers}
    \renewcommand{\arraystretch}{2.2}
    \begin{tabular}{>{\centering\arraybackslash}m{1.5cm}
                    |>{\centering\arraybackslash}m{5cm}
                    >{\centering\arraybackslash}m{5cm}
                    >{\centering\arraybackslash}m{5cm}
                    }
        \hline\hline
        \multirow{2}{*}{\rule[-1ex]{0pt}{2.5ex}~} & \multicolumn{3}{c}{\textbf{Diode-Operated Switches}} \\
        \cline{2-4}
        & \textbf{Conventional} & \textbf{2:1} & \textbf{N:1} \\
        \hline
        $V_{out}$
        & \(\frac{|i_{L}|R_{load}}{\pi}\)
        & \(\frac{|i_{L}|R_{load}}{\pi}(1+\cos(\phi))\)
        & \(\frac{N}{2(N-1)}\left( \frac{|i_L| R_{load}}{\pi}(1+\cos(\phi)) \right)\) \\        
        \hline
        $M$
        & \(\frac{R_{e}}{|Z|}\)
        & \(\left(1+\frac{|Z|}{R^{'}_{e}(\omega)}\right)^{-1}\)
        & \(\left(1+(N-1)\frac{|Z|}{R^{'}_{e}(\omega)}\right)^{-1}\) \\
        \hline
        $\eta$
        & \(\left(1 + \frac{R}{R_{e}}\right)^{-1}\)
        & \(\eta = \left(1 + \frac{R}{R_{e}(\omega)}\right)^{-1}\)
        & \(\eta = \left(1 + \frac{R}{R_{e}(\omega)}\right)^{-1}\) \\
        \hline\hline
        \multirow{2}{*}{\rule[-1ex]{0pt}{0.5ex}~} & \multicolumn{3}{c}{\textbf{Controlled Rectifier}} \\
        \cline{2-4}
        & \textbf{Conventional} & \textbf{2:1} & \textbf{N:1} \\
        \hline
        $V_{out}$
        & \(\frac{|i_{L}|R_{load}}{\pi}\cos(\phi)\)
        & \(2\frac{|i_{L}|R_{load}}{\pi}\cos(\phi)\)
        & \(\frac{N}{(N-1)}\frac{|i_{L}|R_{load}}{\pi}\cos(\phi)\) \\
        \hline
        $M$
        & \(\left(1+\frac{R}{R_e(\omega)}\right)^{-1}\)
        & \(\frac{1}{2}\left(1+\frac{R}{R_e(\omega)}\right)^{-1}\) 
        & \(\frac{1}{N}\left(1+\frac{R}{R_e(\omega)}\right)^{-1}\) \\
        \hline
        $\eta$
        & \(\left(1 + \frac{R}{R_e(\omega)} \right)^{-1}\)
        & \(\left(1 + \frac{R}{R_{e}(\omega)}\right)^{-1}\)
        & \(\left(1 + \frac{R}{R_{e}(\omega)}\right)^{-1}\) \\
        \hline\hline
    \end{tabular}
    \label{tab:tbw}
\end{table*}

\section{Hardware Design and Implementation} 
\begin{figure}[!tb]
\centering
\includegraphics[width=3.0in]{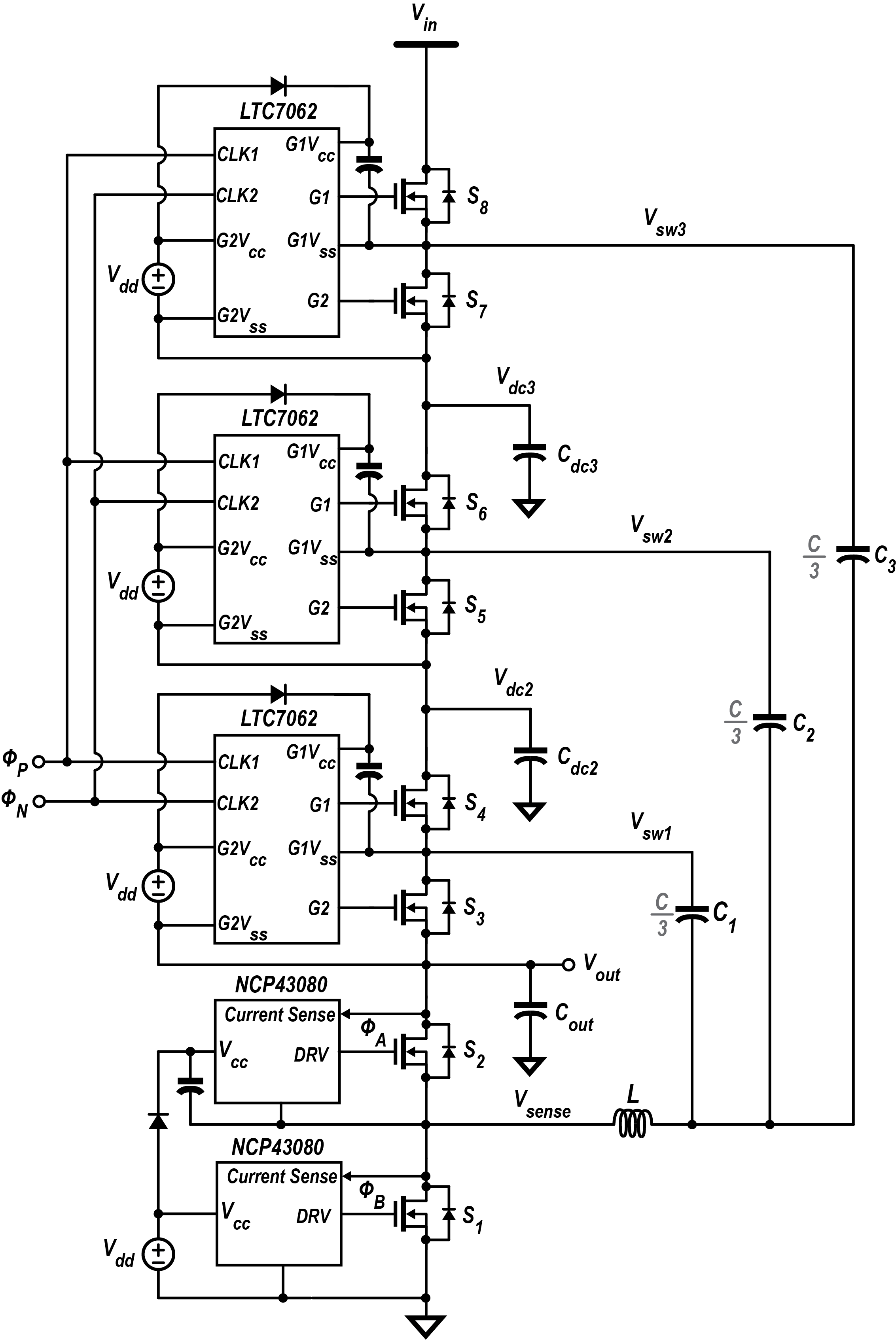}
\caption{Schematic of the implemented 4:1 ReSC converter.}\label{fig.Nx1ReSC_schematic}
\end{figure}
\begin{figure}[!tb]
\centering
\includegraphics[width=2.4in]{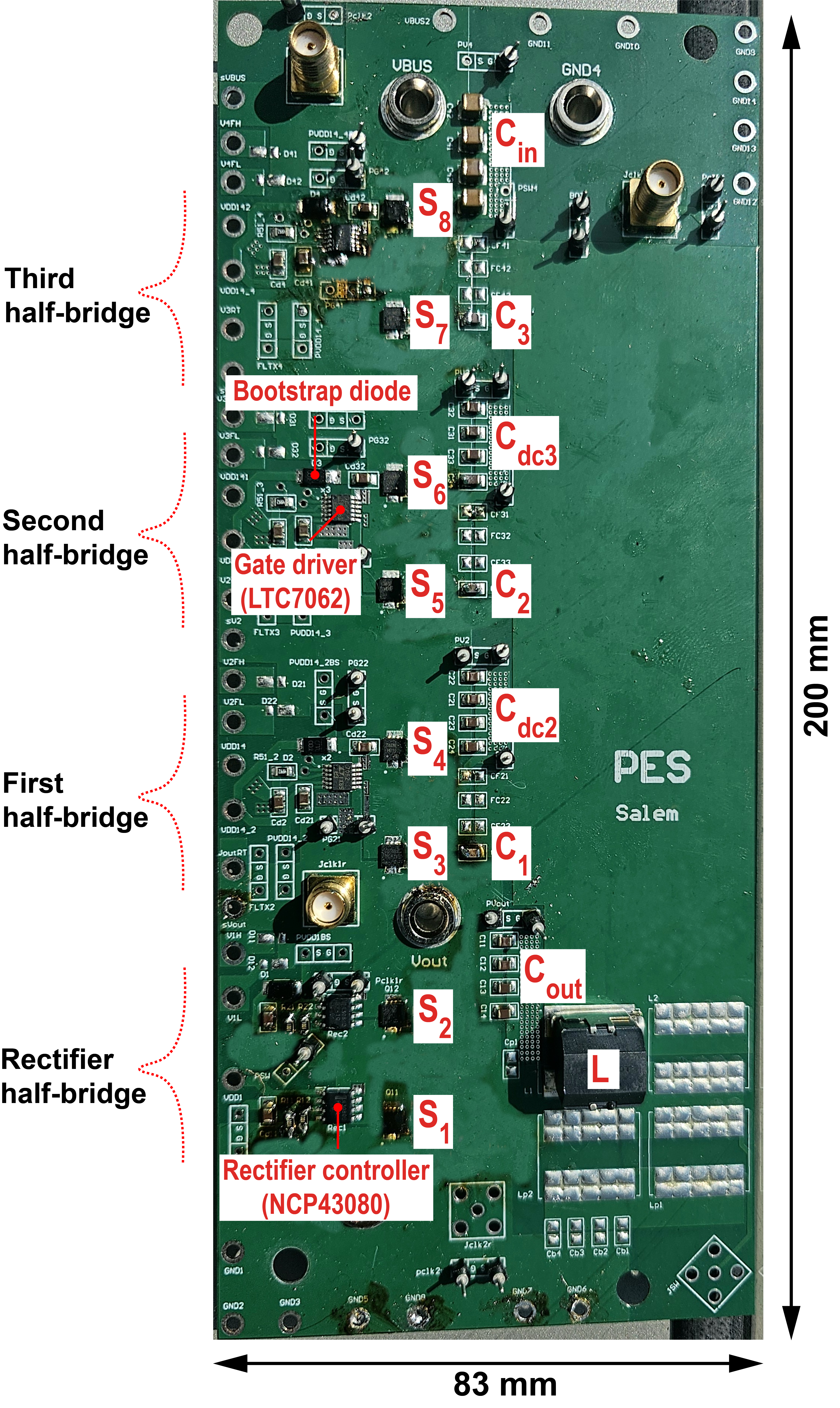}
\caption{Annotated photograph of the experimental 4:1 ReSC converter prototype.}\label{fig.pcb}
\end{figure}

In order to experimentally validate the advantages of the proposed $N$:1 ReSC ladder topology with diode-operated bottom switches, a hardware prototype on a printed circuit board (PCB) is implemented. Specifically, a 4:1 ReSC ladder converter is built that features an input voltage of 40\,V to 60\,V and a constant output voltage of 9\,V. The PCB is a standard, low-cost two-layer (2 oz copper) board with a thickness of 0.6\,mm. The schematic of the implemented 4:1 ReSC converter is shown in Fig.~\ref{fig.Nx1ReSC_schematic}, and the photograph of the hardware prototype PCB, with connectors and additional filter capacitors, is shown in Fig.~\ref{fig.pcb}. The key components are highlighted in the shown figure while their metrics are listed in Table.~\ref{tab:component_list}. A single metal core 1$\,\mu\text{H}$ inductor with 1.06$\,\text{m}\Omega$ DCR is used in the design. The inductor forms with three 33\,nF, 50\,V C0G capacitors the resonant tank of the power converter with a resonant frequency of 506\,kHz. 

 Class II ceramic capacitors (X5R and X7R) offer high energy density that is comparable to electrolytic capacitors~\cite{b188},~\cite{b189}. Therefore, they are used as decoupling capacitors for $V_{in}$, $V_{out}$, and $V_{dc2,3}$ dc nodes that are shown in Fig.~\ref{fig.Nx1ReSC_schematic}. On the other hand, although the flying ladder resonant capacitors $C_{1-3}$ experience equal ac voltage amplitudes, their dc voltage levels increase by integer multiples of $(V_{in}-V_{out})/(N-1)$. The voltage-dependent capacitance reduction characteristic of Class II dielectrics would lead to progressively decreasing capacitance values from $C_{1}$ to $C_{3}$. Instead, class I (C0G) flying capacitors are employed to implement $C_{1-3}$ as they offer better component matching and near-zero voltage coefficients under both dc and ac conditions. They also provide improved loss characteristics and improved stability with aging and temperature~\cite{b187},~\cite{b190}. This is why class I C0G capacitors are typically favored in the implementation of commercial resonant converters~\cite{b191}.

In terms of switch selection, the inherent voltage division capability of the SC ladder topology ensures that each switch only needs to block a fraction of the input voltage, \mbox{$(V_{in}-V_{out})/(N-1)$}, thereby enabling the use of low-voltage MOSFETs rated at 25\,V. Eight identical MOSFETs, $S_{1\text{--}8}$, with an on-resistance of $r_{ON}$\,=\,6\,m$\Omega$ (for $V_{GS}$ = 10\,V and $I_D$ = 20\,A) are used. An NCP43080~\cite{NCP43080} synchronous rectifier controller is used to operate the bottom half-bridge MOSFETs as diodes, enabling optimal power conversion efficiency. The rectifier controller ICs sense the inductor current polarity by monitoring the MOSFETs drain-to-source voltages to generate gate-control signals $\phi_A$ and $\phi_B$, as shown in Fig.~\ref{fig.Nx1ReSC_schematic}. It turns the MOSFET on when the drain-to-source voltage drops below $-$75\,mV, and off when the voltage rises above 0\,mV. The rectifier driver provides a peak source capability of 8\,A and a peak sink capability of 4\,A. This enables fast turn-on (35\,ns) and turn-off (12\,ns) of the MOSFET, ensuring efficient switching performance. Programmable on- and off-time blanking intervals within the IC further contribute to noiseless and reliable operation. The bottom MOSFET rectifier IC is directly powered by a 6\,V supply, while the top MOSFET rectifier IC is powered through a $\sim$6\,V bootstrapped supply referenced to the switching node, $V_{sense}$, as shown in Fig.~\ref{fig.Nx1ReSC_schematic}. A Schottky diode (RSX205LAM30TR) and an X5R capacitor are used to generate the bootstrap supply.

Two low-voltage, ground-referenced out-of-phase external clocks, $\phi_P$ and $\phi_N$ in Fig.~\ref{fig.Nx1ReSC_schematic}, with a controlled dead time are generated using a Tektronix AFG31052 two-channel synchronous signal generator. In this work, an LTC7062 IC~\cite{LTC7062} with a symmetric floating gate driver architecture and a built-in level shifter is used to drive each of the top $N{-}1$ half-bridge inverter switch pairs. The IC provides symmetric gate drivers for both the high-side and low-side MOSFETs of each half bridge. The gate driver of the bottom MOSFET in each stacked half bridge is powered by a 6\,V supply referenced to the half bridge local ground, defined by the static ladder node voltages $V_{out}$, $V_{dc2}$, and $V_{dc3}$. On the other hand, the gate driver of the top MOSFET in each half bridge is powered by a $\sim$\,6\,V bootstrapped supply referenced to the corresponding half-bridge switching node, i.e., $V_{sw1}$, $V_{sw2}$, and $V_{sw3}$. An array of filter capacitors is placed at the static ladder nodes to suppress voltage ripple to a negligible level. While the board layout is designed to minimize parasitic effects, the placement of components is not optimized for compactness, wherein the main purpose of the prototype is to experimentally validate the efficacy of the proposed topology with diode-operated bottom half bridge. The design is scalable to higher conversion ratios by adding additional steps to the ladder. The PCB layout supports this scalability with minimal increase in routing or structural complexity, as shown in Fig.~\ref{fig.pcb}.

\begin{table*}[t]
    \centering
    \caption{Component List of the Hardware Prototype}
       \renewcommand{\arraystretch}{1.5}
    \setlength{\tabcolsep}{30pt}
    \begin{tabular}{c|c|c}
        \hline
        \hline
        \textbf{Component} & \textbf{Part number} & \textbf{Parameters} \\
        \hline
         MOSFET $S_{1-8}$ & Infineon BSZ060NE2LS & 25 V, 6 m$\Omega$ \\
        \hline
         Flying capacitor $C_{1-3}$ & TDK C2012C0G1H333J & C0G, 50 V, 33 nF \\
         Inductor L & TDK PCM120T-1R0N-D & 1 $\mu$H, 1.06 m$\Omega$, 41.2 A \\
        \hline
         Input capacitor $C_{in}$ & Murata GRM32EC72A106KE05K & X7S, 100 V, 10 \textmu F$\times$4 (in parallel)\\
         Output capacitor $C_{out}$ & Murata GRM21BR61E106KA73L & X5R, 25 V, 10 \textmu F$\times$4 (in parallel) \\
         Static ladder capacitor $C_{dc2-dc3}$ & Murata GRM21BR61H106KE43K & X5R, 50 V, 10 \textmu F$\times$4 (in parallel) \\
         \hline
        Gate driver & Analog Devices LTC7062 & Dual high-side gate driver, 100V \\
        Bootstrap diode & Rohm RSX205LAM30TR & Schottky diode, 30V\\
        \hline
        Rectifier controller & Onsemi NCP43080ADR2G & Synchronous rectifier controller, 150V \\
        \hline
        \hline
    \end{tabular}
    
    \label{tab:component_list}
\end{table*}

\begin{table}[h]
    \centering
    \caption{Key Parameters of the Hardware Prototype}
    \renewcommand{\arraystretch}{1.2}
    \setlength{\tabcolsep}{24pt}
    \begin{tabular}{c|c}
        \hline
        \hline
        \textbf{Parameter} & \textbf{Value} \\
        \hline
         Input voltage & 40 - 60 V \\
         Tested output voltage & 9 V \\
         Load resistance & 6 $\Omega$ - 30 $\Omega$ \\
         Switching frequency &  570 kHz - 1.7 MHz \\
         Gate drive voltage & 6 V \\
         Passive component volume\textsuperscript{*} &  0.108 in\textsuperscript{3}\\
         Power component volume\textsuperscript{+} & 0.121 in\textsuperscript{3}\\
         Total volume\textsuperscript{$\dagger$} & 0.127 in\textsuperscript{3}\\
        \hline
        \hline
    \end{tabular}
    \vspace{0.5em}
    \begin{minipage}{1\linewidth}
\justifying
\noindent
\textsuperscript{*}The passive component volume is calculated as the total volume of the flying capacitors and the inductor in the power stage that participate in transferring power to the output, excluding the decoupling capacitors.\\
\textsuperscript{+}The power component volume is defined as the combined volume of all switching devices, capacitors, and inductors in the power stage that contribute to power transfer to the output, excluding the gate drive circuitry.\\
\textsuperscript{$\dagger$}The total volume is estimated as the combined volume of all switching devices, capacitors, and inductors in the power stage that contribute to power transfer to the output, including the gate drive and rectifier ICs.
\end{minipage}
    \label{tab:key_parameter}
\end{table}

\begin{figure}[!tb]
  \centering
  \subfloat[]{\includegraphics[width=3.0in]{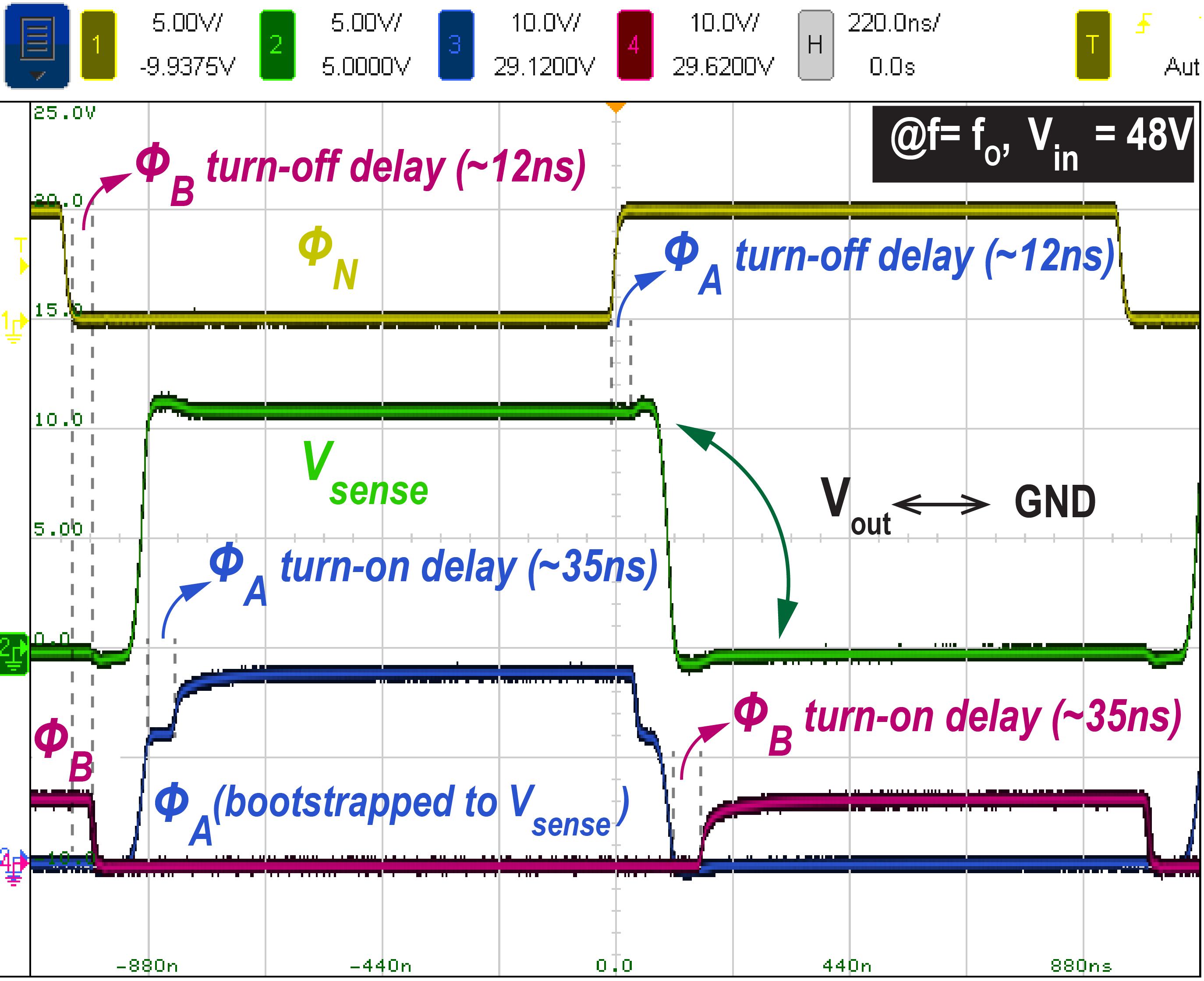}\label{fig.measresults_a}}\\
  \subfloat[]{\includegraphics[width=3.0in]{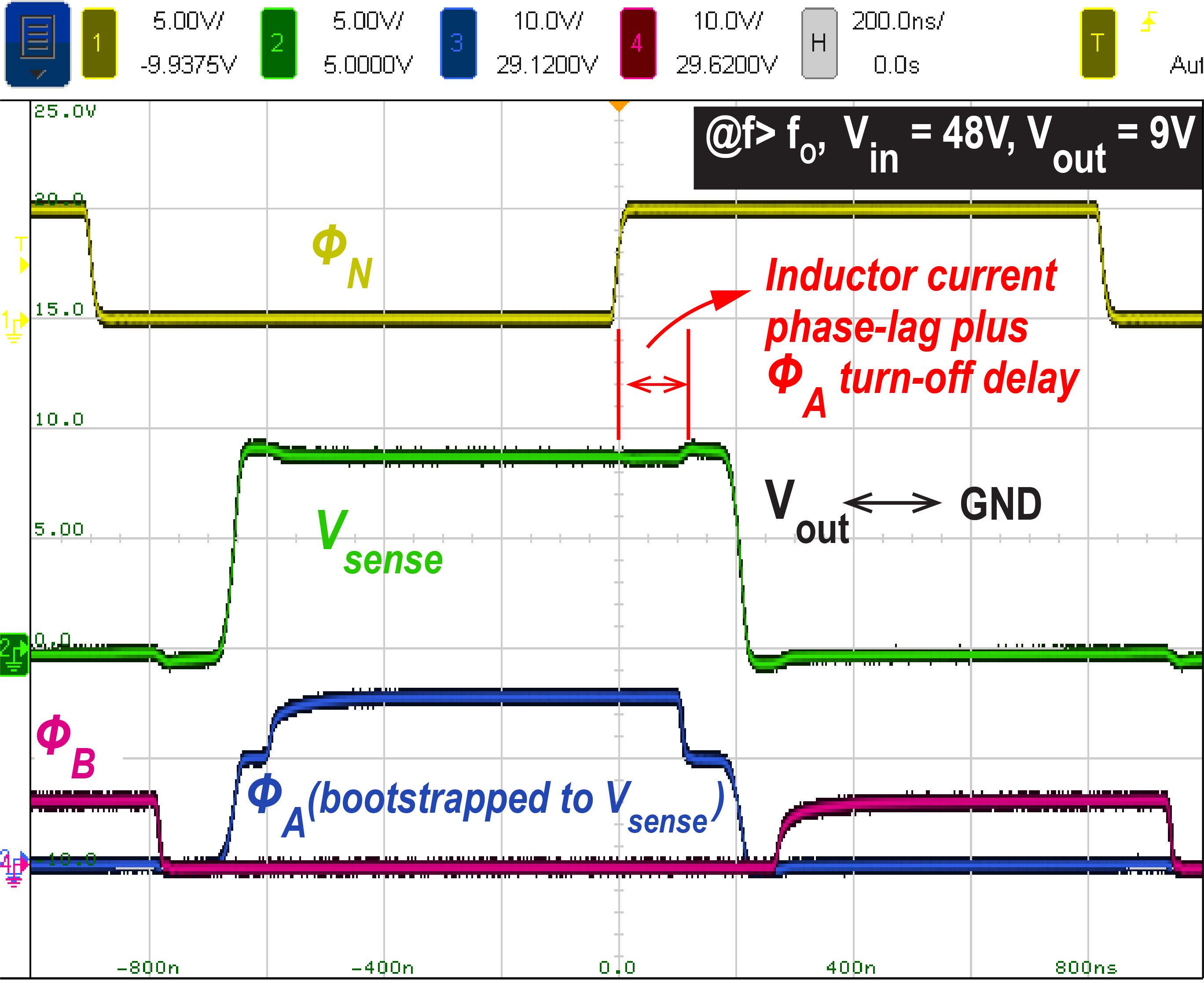}\label{fig.measresults_b}}
  \caption{Measured waveforms of $V_{sense}$ and rectifier clocks, $\phi_{A}$ and $\phi_{B}$, with respect to the top half-bridge clock, $\phi_{N}$, (a) at resonance, $f = f_{o}$, and (b) above resonance, $f > f_{o}$.} \label{fig.measresults}
\end{figure}


\section{Experimental Verification}

Table \ref{tab:key_parameter} summarizes the key parameters of the measured converter. The prototype has been tested with loads as low as 6\,$\Omega$ using a Keysight EL34143A dc electronic load operated in a 4-wire sense mode. A B\&K Precision 5493C 6.5-digit digital multimeter is used to measure the supply current with high precision, enabling accurate efficiency calculations. Fig.~\ref{fig.measresults}(a) shows the external clock signal $\phi_{N}$, the measured rectifier switching node voltage, $V_{sense}$, and the gate control signals, $\phi_{A}$ and $\phi_{B}$, under resonant operation. The rectifier IC’s turn-on delay of approximately 35\,ns and turn-off delay of approximately 12\,ns are annotated. The 35\,ns delay represents only 1.76\% of the switching cycle in worst case, which leads to a negligible impact on the converter efficiency, according to circuit simulation results. Fig.~\ref{fig.measresults}(b) illustrates the converter operation above resonance. The rectifier controller IC (NCP43080) operates switches $S_{1,2}$ as diodes and maintains $V_{sense}$ in phase with the inductor current $i_{L}$ across different switching frequencies $f$. This phase alignment eliminates the imaginary part of $Z^{'}_{e}$ and minimizes the converter conduction loss. 

 \begin{figure}[!tb]
    \centering
    \subfloat[]{%
        \includegraphics[width=2.6in]{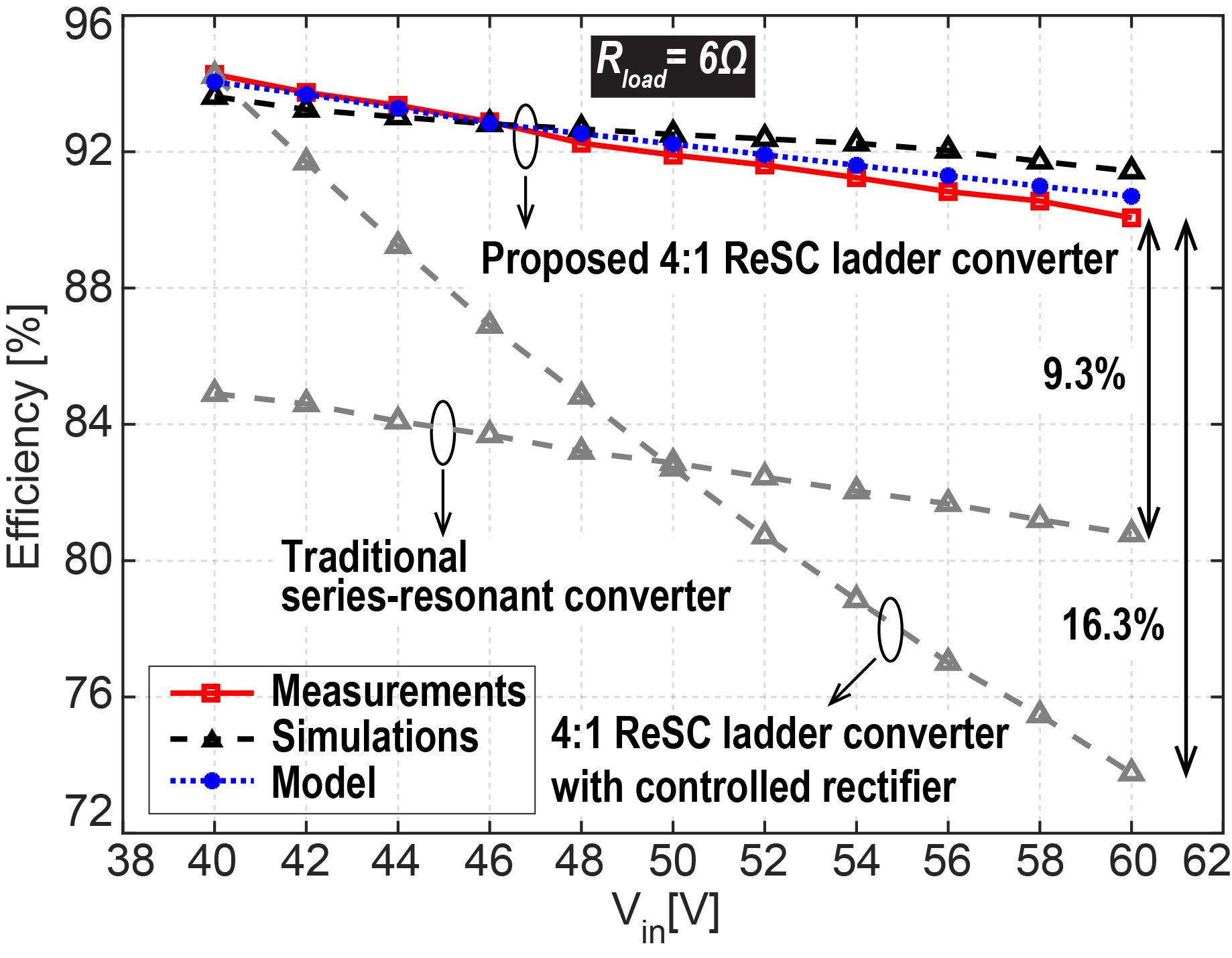}
        \label{fig.meas_eff_6}
    }\\
    \subfloat[]{%
        \includegraphics[width=2.63in]{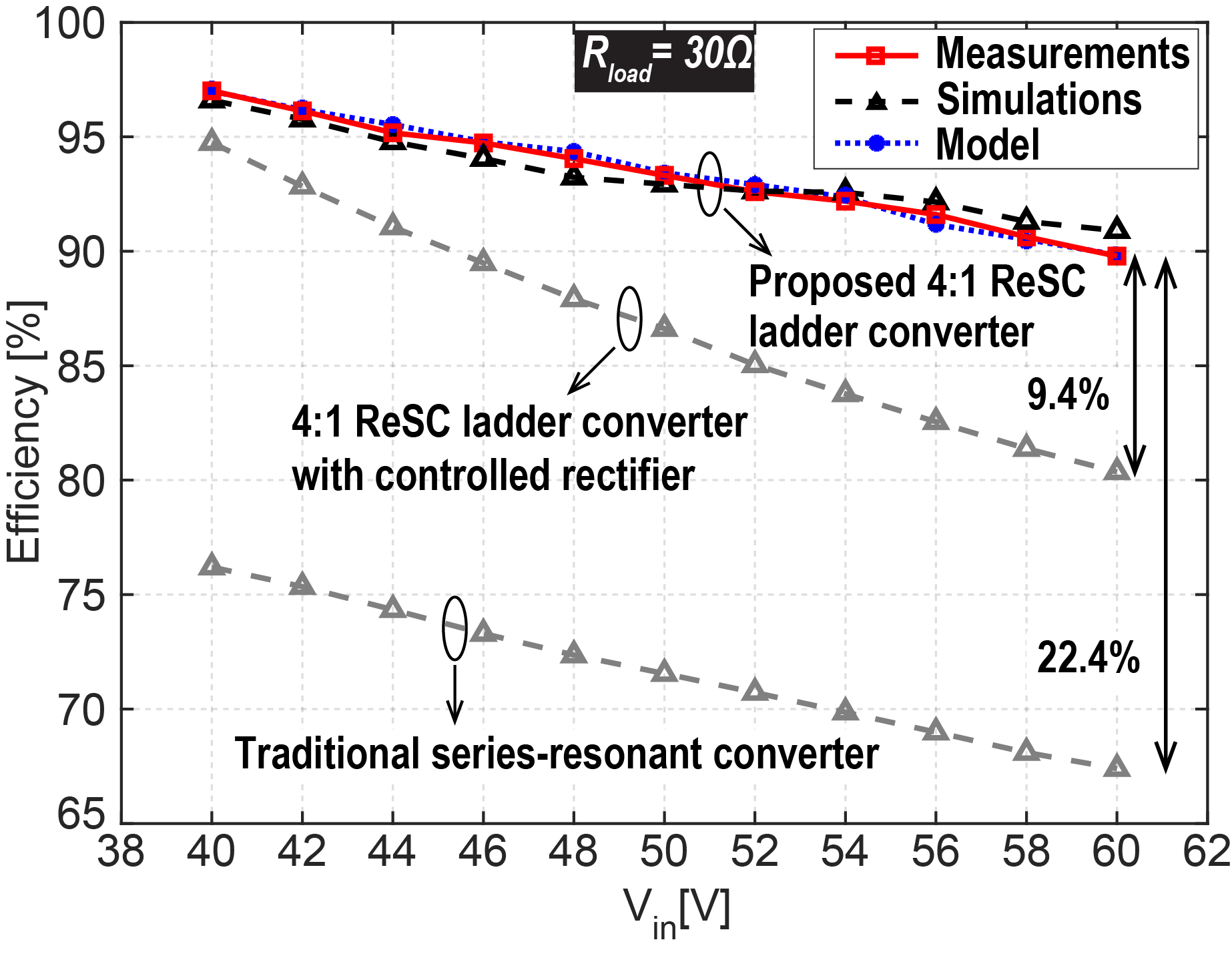}
        \label{fig.meas_eff_30}
    }
    \caption{Measured efficiency of the hardware prototype versus the input voltage under $R_{load}$ of (a) 6$\Omega$ and (b) 30$\Omega$. The simulation results and model \eqref{eq:etaNx1ReSC} predictions are also shown for comparison.}
    \label{fig.meas_eff}
\end{figure}

The measurement results in Fig.~\ref{fig.meas_eff}(a) and (b) show the efficiency of the converter versus the input voltage for an output voltage of 9\,V under a constant load resistance of 6\,$\Omega$ and 30\,$\Omega$, respectively.
The proposed converter achieves a peak efficiency of 97\% under a 30\,$\Omega$ load, while enabling greater than 90\% efficiency over an input voltage range from 40 to 60\,V across both loads, 30\,$\Omega$ and 6\,$\Omega$, illustrating the principal advantage of the proposed diode-operated bottom FETs in realizing efficient wide-range voltage regulation.
The measured efficiency agrees with the theoretical model~\eqref{eq:etaNx1ReSC} predictions within 0.63\%, as shown in Fig.~\ref{fig.meas_eff}. The model includes an estimated inductor ESR of (1.06 + 2.385\,\textmu $\times$$f^{1.23}$)\,\text{m}$\Omega$, 13\% of the MOSFET switching losses modeled as $C_{OSS}V^2f$, where $C_{OSS}$ is the transistor output capacitance, and an additional resistance of 18\,m$\Omega$ to account for the increase in MOSFET on-resistance and inductor ESR due to temperature rise as well as PCB parasitics. In metal composite cores, eddy current losses typically scale with frequency between $f^{1.5}$ and $f^2$ ~\cite{racmodelref1, racmodelref2}. For this work, an exponent of $f^{1.23}$ is used to best fit the frequency-dependent ac resistance model to the manufacturer’s data over the employed operating range~\cite{PCM120T}. 
The proposed ladder ReSC topology achieves up to 16.3\% and 9.4\% higher efficiency compared to a ladder ReSC topology with a controlled rectifier, and up to 9.3\% and 22.4\% higher efficiency compared to a traditional series-resonant converter, as shown in Fig.~\ref{fig.meas_eff}(a) and (b), under $R_{\text{load}}\,=\,6\,\Omega$ and $30\,\Omega$, respectively. The traditional series-resonant converter employs ten BSZ060NE2LS MOSFETs, similar to the ones used for the implemented prototype. 
Four FETs are connected in series to realize each switch, $S_{3}$ and $S_{4}$, in the half-bridge inverter shown in Fig. \ref{fig.conv1xresconv}(a). The efficiency improvement over the conventional resonant converter is attributed to the ladder topology that effectively connects the FETs in parallel, in the ac sense, reducing their on-resistance to $r_{ON}/3$, as illustrated in Fig.~\ref{fig.Nx1ReSC}(b), whereas the traditional series-resonant topology connects the switches in series, resulting in a higher inverter resistance of 4$r_{ON}$.

\begin{table*}[!th]
\fontsize{7pt}{7pt}
\caption{\textcolor{black}{Comparison Of The Proposed 5:1 Ratio Resonant Switched-Capacitor DC-DC Converter With The Prior Art}\label{tab:comparison_table_iscas}}
\centering
\renewcommand{\arraystretch}{1.0}
\setlength{\tabcolsep}{0.1pt}
\begin{tabular}{|c|c|c|c|c|c|c|c|c|}
       \hline
\textcolor{black}{\textbf {Design}} &\textcolor{black}{\textbf{C Schaef}} &\textcolor{black}{\textbf{P McLaughlin}} &\textcolor{black}{\textbf{Tingxu Hu}} &\textcolor{black}{\textbf{W L Zeng}} &\textcolor{black}{\textbf{Xu Yang}} &\textcolor{black}{\textbf{G Cai}} &\textcolor{black}{\textbf{Tingxu Hu}} &\textcolor{black}{\textbf{This work}}\\

&\textcolor{black}{\textbf{JESTPE’18}\textbf{\cite{b0}}} 
& \textcolor{black}{\textbf{ISSCC’22} \textbf{\cite{b1}}} 
& \textcolor{black}{\textbf{ISSCC’22} \textbf{\cite{b2}}} 
& \textcolor{black}{\textbf{ISSCC’23} \textbf{\cite{b3}}} 
& \textcolor{black}{\textbf{ISSCC’23} \textbf{\cite{b4}}} 
& \textcolor{black}{\textbf{ISSCC’23} \textbf{\cite{b5}}} 
& \textcolor{black}{\textbf{ISSCC’23} \textbf{\cite{b24}}} 
& \\

\hline
\textbf{\textcolor{black}{Technology}} 
& \textcolor{black}{180nm} 
& \textcolor{black}{180nm} 
& \textcolor{black}{180nm HV SOI} 
& \textcolor{black}{180nm BCD} 
& \textcolor{black}{180nm BCD} 
& \textcolor{black}{180nm BCD} 
& \textcolor{black}{180nm BCD} 
& \textbf{\textcolor{black}{180nm}} \\

\hline
\textbf{\textcolor{black}{Topology}} 
& \textcolor{black}{3:1 ReSC} 
& \textcolor{black}{3:1 ReSC} 
& \textcolor{black}{CCC} 
& \textcolor{black}{2L4PHB} 
& \textcolor{black}{Dual path- SC} 
& \textcolor{black}{ReSC-PL} 
& \textcolor{black}{SC Buck} 
& \textbf{\textcolor{black}{5:1 ReSC}} \\
\hline
\textbf{\textcolor{black}{Input voltage [V]}} 
& \textcolor{black}{12} 
& \textcolor{black}{3.3-6.6} 
& \textcolor{black}{12} 
& \textcolor{black}{12} 
& \textcolor{black}{9-16} 
& \textcolor{black}{12} 
& \textcolor{black}{12} 
& \textbf{\textcolor{black}{9}} \\
\hline
\textbf{\textcolor{black}{Output voltage [V]}} 
& \textcolor{black}{3.7} 
& \textcolor{black}{0.8-2.2} 
& \textcolor{black}{0.9-1.8} 
& \textcolor{black}{1-1.8} 
& \textcolor{black}{0.6-1.6} 
& \textcolor{black}{0.6-1.2} 
& \textcolor{black}{0.8-1.8} 
& \textbf{\textcolor{black}{0.8-1.5}} \\
\hline
\textbf{\textcolor{black}{Max load current [A]}} 
& \textcolor{black}{1.25} 
& \textcolor{black}{0.48} 
& \textcolor{black}{4} 
& \textcolor{black}{4} 
& \textcolor{black}{5} 
& \textcolor{black}{5} 
& \textcolor{black}{4*} 
& \textbf{\textcolor{black}{3.3}} \\
\hline
\textbf{\textcolor{black}{Max output power [W]}} 
& \textcolor{black}{4.62} 
& \textcolor{black}{0.67} 
& \textcolor{black}{7.2} 
& \textcolor{black}{4.5} 
& \textcolor{black}{8} 
& \textcolor{black}{6} 
& \textcolor{black}{7.2} 
& \textbf{\textcolor{black}{4.95}} \\

\hline
 & \textcolor{black}{36nH} 
 & \textcolor{black}{10nH} 
 & \textcolor{black}{2x0.56\textmu H} 
 & \textcolor{black}{2x1\textmu H} 
 & \textcolor{black}{0.68\textmu H} 
 & \textcolor{black}{1\textmu H} 
 & \textcolor{black}{220nH} 
 & \textbf{\textcolor{black}{240nH (0603,19m$\Omega$)}} \\
\textbf{\textcolor{black}{Passive elements}} 
 & \textcolor{black}{(0402,34m$\Omega$)} 
 & \textcolor{black}{(on-chip)} 
 & \textcolor{black}{(N.R, 14.5m$\Omega$)}  
 & \textcolor{black}{(1008,48m$\Omega$)} 
 & \textcolor{black}{(N.R, N.R)}  
 & \textcolor{black}{(1008,N.R)} 
 & \textcolor{black}{(1210, 6m$\Omega$)} 
 & \textbf{\textcolor{black}{4x25nF (single 0603)}} \\
\textbf{\textcolor{black}{for power conversion}} 
 & \textcolor{black}{2x0.47\textmu F} 
 & \textcolor{black}{2x2.7nF} 
 & \textcolor{black}{2x2.2\textmu F} 
 & \textcolor{black}{3x22\textmu F} 
 & \textcolor{black}{4x10\textmu F} 
 & \textcolor{black}{2x22\textmu F (0805)} 
 & \textcolor{black}{4*4.7\textmu F}  
 & \textbf{\textcolor{black}{3x4.7\textmu F (0201)}} \\
 & \textcolor{black}{(0402)}  
 & \textcolor{black}{(on-chip)} 
 & \textcolor{black}{(N.R)} 
 & \textcolor{black}{(N.R)}  
 & \textcolor{black}{(N.R)} 
 & \textcolor{black}{2x10\textmu F (0603)} 
 & \textcolor{black}{(0805)} 
 & \textbf{\textcolor{black}{2x10nF (0201)}} \\
\hline
\textbf{\textcolor{black}{Output decoupling}} 
 & \textcolor{black}{4.7\textmu F} 
 & \textcolor{black}{5.8nF} 
 & \textcolor{black}{10.6\textmu F} 
 & \textcolor{black}{22\textmu F} 
 & \textcolor{black}{22\textmu F} 
 & \textcolor{black}{100\textmu F (0805)} 
 & \textcolor{black}{2x22\textmu F} 
 & \textbf{\textcolor{black}{2x4.7\textmu F}} \\
\textbf{\textcolor{black}{capacitor ($C_{out}$)}} 
 & \textcolor{black}{(0402)}  
 & \textcolor{black}{} 
 & \textcolor{black}{(N.R)} 
 & \textcolor{black}{(N.R)} 
 & \textcolor{black}{(N.R)} 
 & \textcolor{black}{10nF (0402)} 
 & \textcolor{black}{(0805)} 
 & \textbf{\textcolor{black}{(0201)}} \\
\hline
\textbf{\textcolor{black}{Input decoupling}} 
 & \textcolor{black}{0.22\textmu F} 
 & \textcolor{black}{4.2nF} 
 & \textcolor{black}{N.R} 
 & \textcolor{black}{N.R} 
 & \textcolor{black}{N.R} 
 & \textcolor{black}{N.R (0805)**} 
 & \textcolor{black}{2$\times$4.7\textmu F} 
 & \textbf{\textcolor{black}{4.7\textmu F}} \\
\textbf{\textcolor{black}{capacitor ($C_{in}$)}} 
 & \textcolor{black}{(0201)} 
 & \textcolor{black}{} 
 & \textcolor{black}{} 
 & \textcolor{black}{} 
 & \textcolor{black}{} 
 & \textcolor{black}{} 
 & \textcolor{black}{(0805)} 
 & \textbf{\textcolor{black}{(0201)}} \\
\hline
\textbf{\textcolor{black}{Total volume(V)[mm$^{3}$]†}} 
 & \textcolor{black}{10.05} 
 & \textcolor{black}{7.40} 
 & \textcolor{black}{32.43***} 
 & \textcolor{black}{28.15***} 
 & \textcolor{black}{50.58***} 
 & \textcolor{black}{21.85} 
 & \textcolor{black}{21.73} 
 & \textbf{\textcolor{black}{4.84}} \\
\hline
\textbf{\textcolor{black}{Peak efficiency (PE) ($\eta$\textsubscript{PE}) [\%]}} 
 & \textcolor{black}{87.00} 
 & \textcolor{black}{78.30} 
 & \textcolor{black}{91.00} 
 & \textcolor{black}{93.70} 
 & \textcolor{black}{94.50} 
 & \textcolor{black}{91.80} 
 & \textcolor{black}{90.00} 
 & \textbf{\textcolor{black}{96.56}} \\
\hline
\textbf{\textcolor{black}{Power density at PE [W/mm$^{3}$]}} 
 & \textcolor{black}{0.46} 
 & \textcolor{black}{0.02} 
 & \textcolor{black}{0.08} 
 & \textcolor{black}{0.12} 
 & \textcolor{black}{0.02} 
 & \textcolor{black}{0.05} 
 & \textcolor{black}{0.08} 
 & \textbf{\textcolor{black}{0.08}} \\
\hline
\textbf{\textcolor{black}{Peak power density (PPD)††}} 
 & \textcolor{black}{0.46} 
 & \textcolor{black}{0.09} 
 & \textcolor{black}{0.22} 
 & \textcolor{black}{0.16} 
 & \textcolor{black}{0.16} 
 & \textcolor{black}{0.27} 
 & \textcolor{black}{0.33} 
 & \textbf{\textcolor{black}{1.02}} \\
\hline
\textbf{\textcolor{black}{Efficiency at PPD [\%]}} 
 & \textcolor{black}{87.00} 
 & \textcolor{black}{61.00} 
 & \textcolor{black}{84.90} 
 & \textcolor{black}{91.00} 
 & \textcolor{black}{84.40} 
 & \textcolor{black}{81.00} 
 & \textcolor{black}{83.80} 
 & \textbf{\textcolor{black}{85.20}} \\
\hline

\end{tabular}
\\
\vspace{0.1cm}
\textcolor{black}{*Single-output, **Estimated from the Printed Circuit Board, ***Estimated from the paper, †PMIC (0.85mm chip) + passives + $C_{out}$ + $C_{in}$ volume, \\
††Peak power density (PPD)=Max output power/Total volume [W/mm$^{3}$]  N.R = not reported.}
\end{table*}

\textcolor{black}{\section{Comparison to Existing Methods for High Step-Down Voltage Conversion}}
\begin{figure}[!tb]
\centering
\includegraphics[width=3.4in]{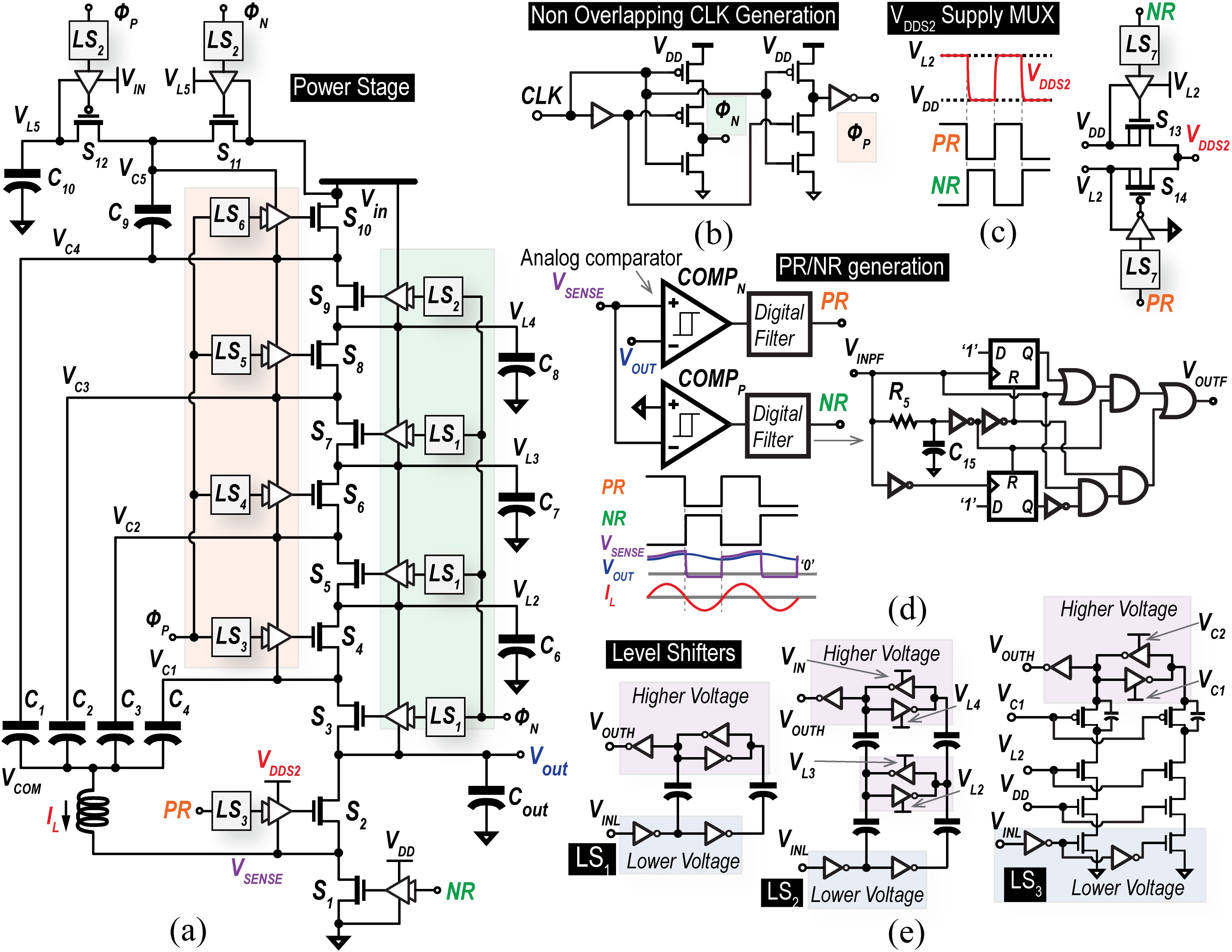}
\vspace*{-0.4cm} 
\caption{Circuit schematics of the (a) power stage; (b) non-overlapping clock generation; (c) $V_{DDS2}$ supply MUX; (d) PR and NR control generation; and (e) level-shifters $LS_{1-3}$.}
\label{fig.circuitsch}
\end{figure}

\begin{figure}[!tb]
\centering
\includegraphics[width=2.15in]{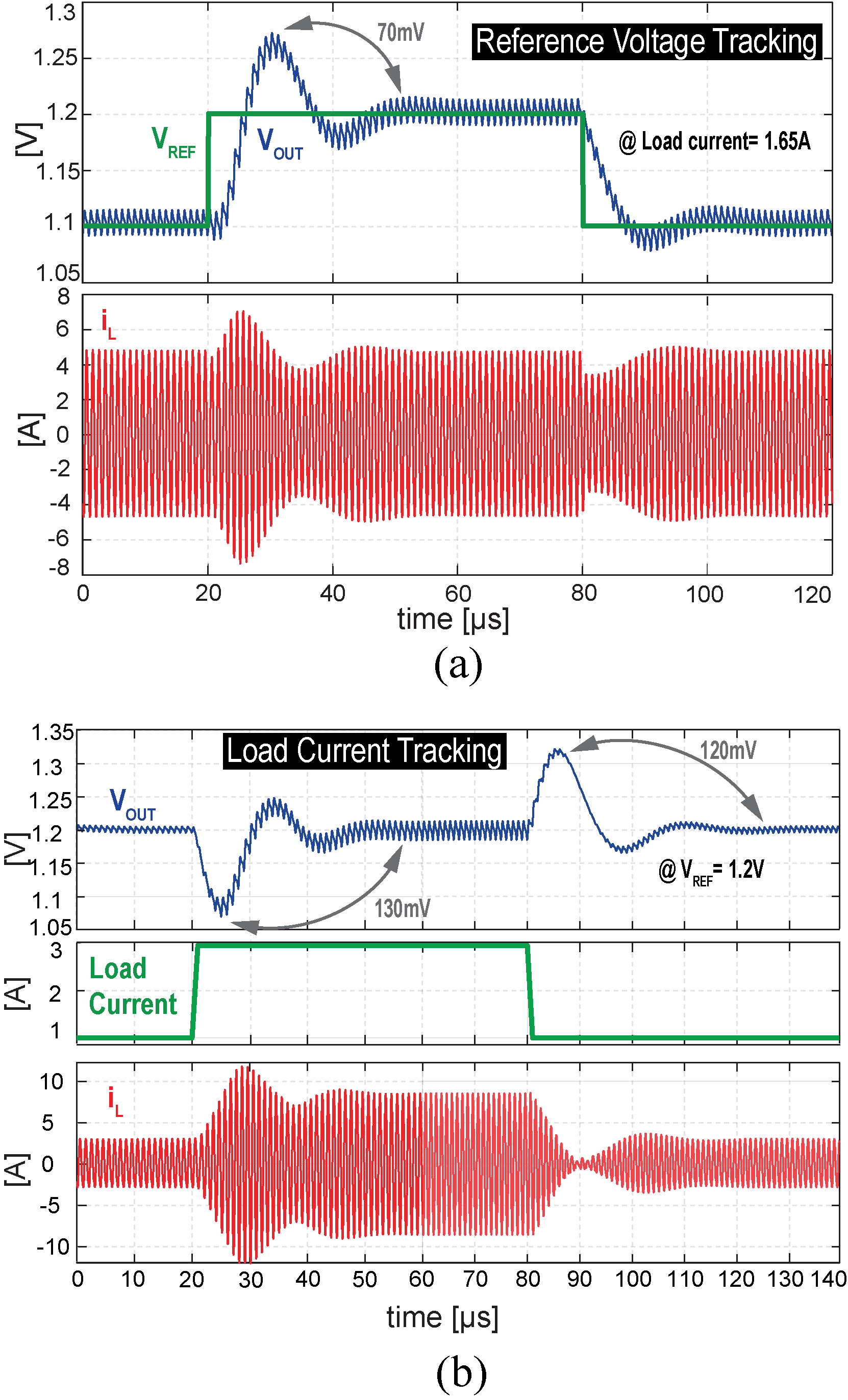}
\caption{(a) Post-layout simulated reference voltage tracking waveforms. (b) Post-layout simulated load current tracking waveforms. }
\label{fig.Controller_transients}
\end{figure}

\textcolor{black}{This section demonstrates the advantages of the proposed topology with diode-operated bottom switches in realizing steep voltage conversion compared to existing techniques. A proof-of-concept 5:1 ReSC converter of the proposed topology is designed using 0.18\textmu m CMOS technology and its performance is compared to existing approaches for steep voltage conversion. The schematic of the proof-of-concept and implementation details are discussed in \cite{ISCAS_conf}.}
\textcolor{black}{Ten thin-oxide NMOS transistors ($S_{1-10}$) form the power stage\textcolor{black}{, as illustrated in Fig. \ref{fig.circuitsch}}. These transistors are driven by two non-overlapping clock phases, $\phi_P$ and $\phi_N$, that are provided from a controller. 
On the other hand, rectifier clock phases, PR and NR, for the bottom half-bridge are generated using two analog comparators, which compare the half-bridge switching node voltage, $V_{sense}$, with $V_{out}$ and ground. 
The 0.18\textmu m proof-of-concept converter varies the duty cycle (D) of the clock signal provided to the top eight transistors while operating the bottom transistors as diodes to facilitate lossless $V_{out}$ regulation. 
A type-III compensator is implemented to ensure stable closed-loop operation \cite{ISCAS_conf}.} 

\textcolor{black}{The converter is evaluated with a 9 V input supply. A 240 nH inductor with 19 m$\Omega$ DC resistance and four 25 nF capacitors form the resonant tank, yielding a natural frequency of 1.03 MHz. Efficiency is characterized across load currents from 50 mA to 3.3 A and output voltages between 0.8–1.5 V. The simulated peak efficiency reaches 96.56\% at 0.25 A load, while 85.2\% efficiency is achieved at the maximum load of 3.3 A with a 1.5 V output. In closed-loop operation, the converter tracks the reference voltage with a steady-state error of –0.5\% to 1.5\%. During reference steps between 1.1 V and 1.2 V at a 1.65 A load, the output is regulated within 70 mV overshoot and 20 mV undershoot \textcolor{black}{, as shown in Fig.\ref{fig.Controller_transients}(a)}. For a load transient from 1 A to 3 A within 0.1\textmu s (and vice versa), the simulated undershoot and overshoot remain below 130 mV and 120 mV, respectively \textcolor{black}{, as shown in Fig.\ref{fig.Controller_transients}(b)}.}
\textcolor{black}{Table \ref{tab:comparison_table_iscas} presents a comparison of the proposed resonant converter against prior hybrid and resonant designs with comparable input and output voltage ranges. The 5:1 resonant converter demonstrates superior performance, achieving the highest peak efficiency and power density among existing solutions.}

\section{Conclusion}
This paper presented a single-inductor $N$:1 resonant switched-capacitor ladder converter with continuous voltage conversion capability. The use of diode-operated switches in the bottom half-bridge enables automatic and optimal phase-shift tracking of the inductor current relative to the half-bridge switching clock. A theoretical model was developed to predict the conversion ratio as well as efficiency of the converter, and its accuracy was validated through circuit simulations for $N$ = 1, 2, and 4 as well as measurement results. A hardware prototype was implemented to demonstrate the advantages of the proposed solution for 48\,V to 9\,V conversion. 
The measured efficiency closely matches the model predictions, with deviations of less than 0.53\% and 0.63\% under load resistances of 30\,$\Omega$ and 6\,$\Omega$, respectively.




\vfill

\end{document}